\documentclass{article}

\usepackage[margin=1 in]{geometry}

\usepackage{amsmath}
\usepackage{amssymb}
\usepackage{mathabx}
\usepackage{graphicx}
\usepackage{dcolumn}
\usepackage{bm}
\usepackage{siunitx}
\usepackage{enumitem}
\usepackage{xcolor}
\usepackage{float}
\usepackage{comment}
\usepackage{chemformula}
\usepackage[makeroom]{cancel}
\usepackage[numbers,sort&compress]{natbib}
\usepackage{array}
\newcolumntype{P}[1]{>{\centering\arraybackslash}p{#1}}
\title{A Quantitative Analysis of the Ignition Characteristics\\of Fine Iron Particles}

\author
{XiaoCheng Mi$^{\ast,1,2}$, Aki Fujinawa$^{1}$, and Jeffrey M. Bergthorson$^{1}$\\ \\
\normalsize{$^{1}$Department of Mechanical Engineering, McGill University,}\\ 
\normalsize{Montreal H3A 0C3, Canada}\\

\normalsize{$^{2}$Department of Mechanical Engineering, Eindhoven University of Technology,}\\ 
\normalsize{Eindhoven 5600 MB, the Netherlands}\\

\normalsize{$^\ast$Correspondence to: x.c.mi@tue.nl; xiaocheng.mi@mail.mcgill.ca}}

\date{}

\begin{document}
\maketitle

\begin{abstract}
Ignition of iron particles in an oxidizing environment marks the onset of self-sustained combustion. The objective of the current study is to quantitatively examine the ignition characteristics of fine iron particles (i.e., \SI{1}{\micro\meter}- to \SI{100}{\micro\meter}-sized) governed by the kinetics of solid-phase iron oxidation. The oxidation rates are inversely proportional to the thickness of the oxide layer (i.e., following a parabolic rate law) and calibrated using the experimentally measured growth of iron-oxide layers over time. Steady-state (i.e., Semenov's analysis) and unsteady analysis have been performed to probe the dependence of the critical gas temperature required to trigger a thermal runaway (namely, the ignition temperature $T_\mathrm{ign}$) on particle size, initial thickness of oxide layer, inert gas species, radiative heat loss, and the collective heating effect in a suspension of particles. Both analyses indicate that $T_\mathrm{ign}$ depends on $\delta_0$, i.e., the ratio between the initial oxide layer thickness and particle size, regardless of the absolute size of the particle. The unsteady analysis predicts that, for $\delta_0 \lesssim 0.003$, $T_\mathrm{ign}$ becomes independent of $\delta_0$. Under standard conditions in air, $T_\mathrm{ign}$ is approximately \SI{1080}{\kelvin} for any particle size greater than \SI{5}{\micro\metre}. The ignition temperature decreases as the thermal conductivity of the oxidizing gas decreases. Radiative heat loss has a minor effect on $T_\mathrm{ign}$. The collective effect of a suspension of iron particles in reducing $T_\mathrm{ign}$ is demonstrated. The transition behavior between kinetic-controlled and external-diffusion-controlled combustion regimes of an ignited iron particle is systematically examined. The influences of initial oxide-layer thickness and particle temperature on the ignition delay time, $\tau_\mathrm{ign}$, of iron particles are parametrically probed. A $d^2$-law scaling between $\tau_\mathrm{ign}$ and particle size is identified. Possible sources of inaccuracy are discussed.

\end{abstract}

\providecommand{\keywords}[1]
{
  \small	
  \textbf{\textit{Keywords--}} #1
}

\section{\label{Sect1}Introduction}
Iron is an excellent fuel for long-term storage and long-distance transport of clean energy owing to its carbon-free nature, high energy density, and potential for non-volatile combustion in air~\cite{Beach2006,Bergthroson2015Review,Bergthorson2018Review}. For developing practical energy-conversion technologies based on iron fuel, a better understanding of the fundamentals underlying the combustion process of fine (i.e., micron- to hundreds-of-micron-sized) iron particles at elevated temperatures is required. The full oxidation process of an iron particle in a combustion system consists of three major stages: (1) Preheating and ignition leading to a thermal runaway; (2) a rapid oxidation process of a molten droplet at elevated temperatures until iron is completely oxidized; (3) a slow further oxidation process of the lower oxidized products, i.e., from \ch{FeO} or \ch{Fe3O4} to \ch{Fe2O3}, upon cooling. To design a real-world combustion device of iron particles, quantitative answers are required for the following questions: (A) Under what conditions can iron particles be ignited? (B) How to ensure a non-volatile combustion of molten iron droplets? (C) What are the rates of energy release during the rapid and slow iron oxidation processes? To answer these questions, we need to better identify the rate-controlling mechanisms among a complex of physicochemical processes in the different stages of iron combustion, including oxygen diffusion in the gas, oxygen absorption at the particle surface, phase change and dissociative evaporation of iron and its oxides, solid- and liquid-phase kinetics of iron oxidation, and heat exchange between particles and the ambient gas. Both experimental and theoretical endeavors are required for this exploration.\\

Prior to the research campaigns on fine iron particles, knowledge of iron combustion has mostly been acquired from experimental investigations on the growth of solid iron-oxide scales in air~\cite{Paidassi1958a,Paidassi1958b,Yurek1974,SmeltzerYoung1975,GarnaudRapp1977} and iron rod combustion in high pressure oxygen~\cite{Sato1983,Hirano1993,Steinberg1992,Steinberg1998,WardSteinberg2009,Muller2014CST,Muller2015}. Metallurgical studies~\cite{Paidassi1958a,Paidassi1958b} dated in the 1950s demonstrated that the growth of solid-phase iron oxides on a 
specimen of pure iron at a steady temperature follows a parabolic rate law, namely, with a rate inversely proportional to the thickness of the oxide scale. This finding agrees well with the prediction of Wagner's theory~\cite{Wagner1933} asserting that the rate of solid-phase metal oxidation is controlled by the lattice diffusion of ions across the oxide scale~\cite{Hauffe1965}. Quantitatively accurate kinetics for solid-phase iron oxidation can be determined from these experimental results.\\

The combustion complex of a \SI{}{\milli\meter}-diameter iron rod consists of a molten droplet of iron and iron oxides attached to a reaction front that melts and propagates into the solid rod~\cite{Sato1983}. Dreizin~\cite{Dreizin2000} elucidated that the burning rate of an iron rod can be either limited by the iron oxidation kinetics at sufficiently high gas temperatures~\cite{Steinberg1992,Steinberg1998} or the incorporation rate of gaseous oxygen at low gas temperatures~\cite{Sato1983,Hirano1983}. Recent studies by Muller \textit{et al}.~\cite{Muller2014CST,Muller2015} reveal more details of the molten mixture formed on the top of a laser-ignited iron rod. A spatially non-uniform distribution of molten iron, oxides, and gas bubbles inside the droplet was identified~\cite{Muller2015}. A subsequent modeling work~\cite{ElRabii2017} further describes the complex thermo-fluid processes within this molten-phase mixture controlling the burning rate of an iron rod. Alas, quantitatively useful information to answer the aforeposed questions for iron-particle combustion is mostly masked by the intrinsic complexity of iron-rod combustion.\\

 Over the past two decades, the state-of-the-art understanding in the combustion of fine iron particles has been advanced by researchers around the globe. This collective effort has, however, been mainly focused on experimentally examining~\cite{SunHirano2000,Tang2011PROCI,Julien2015,Wright2016,McRae2018,Palecka2019Perwaves,Palecka2020Perwaves,Poletaev2020} and qualitatively modeling~\cite{Tang2009CTM,Tang2011PRE,Goroshin2011PRE,Mi2016PROCI,Lam2017PRE,Lam2018CTM} the macroscopic observables of flame propagation in a dispersion of iron particles in an oxidizing gas. Although a few recent experimental studies~\cite{Toth2020,Ning2021,Huang2021,Ning2022CNF} attempt to monitor the combustion process of individual iron particles, \textit{in situ} measurements of particle temperature and composition have not yet been \textit{extensively} obtained. Thus, the detailed rate-controlling mechanisms of iron-particle combustion, especially for a molten droplet, remain elusive. Without a physics-based, quantitatively reliable model for calculating the energy release rate of an iron particle, the development of high-fidelity simulations of large-scale combustion systems is baseless despite the sophistication achieved in modeling other components of the problem.\\
 
 This paper reports an effort in seeking a quantitative answer to the aforementioned Question (A) with regard to the \textit{ignition characteristics} of fine iron particles. To ignite an iron particle means that a set of critical conditions (e.g., gas or particle temperature) is satisfied to trigger a thermal runaway---the rate of energy release due to iron oxidation exceeds the rate of heat loss to the surrounding. There is yet no experimental measurement of the critical particle or gas temperature required for triggering thermal runaway (referred to as ``ignition temperature’’ in the remainder of this paper) of an isolated particle, or a suspension of iron particles, in a flame. The experimental data of iron ignition temperature in the literature were obtained for suspensions and precipitated beds of fine iron powders~\cite{Gorokhov1964,Leshchevich2012} and \SI{}{\milli\meter}- to \SI{}{\centi\meter}-sized iron or steel specimens~\cite{Grosse1958,Bolobov1991,Bolobov2001}. These data, as partly summarized by Breiter~\textit{et al}.~\cite{Breiter1977}, are widely scattered owing to the difference in experimental conditions, specimen properties and morphologies, and definition of ``ignition temperature''. The experimental measurement reported by Grosse and Conway~\cite{Grosse1958} and Bolobov~\cite{Bolobov2001} are perhaps the closest to the herein defined ignition temperature. In Grosse and Conway’s study~\cite{Grosse1958}, a 10-\SI{}{\gram}-weighted specimen of iron was placed in an Alundum crucible and heated to a designated temperature in an argon atmosphere by an electric furnace, and then, a flow of oxygen was fed to the top of the crucible. This procedure was repeated with an incremented temperature until an event of thermal runaway was detected by visual observation. An ignition temperature of $1203 \pm 10\mathrm{K}$ was obtained. A similar procedure was followed by Bolobov~\cite{Bolobov2001} to determine the ignition temperature of a $5 \times 5 \times \SI{0.5}{\milli\meter}$ low-carbon steel foil. The specimens were heated in vacuum to a designated temperature before supplying oxygen. The event of a thermal runaway was detected as an abrupt increase in the temporal record of specimen temperature.  An ignition temperature of  $1233 \pm 20 \mathrm{K}$, independent of ambient oxygen concentration, was found. Both of these experimental results of ignition temperature are significantly below the melting point of \ch{FeO}, i.e., \SI{1650}{\kelvin}, indicating that the ignition process occurs when a particle remains completely in solid phase.\footnote{The melting points of \ch{Fe}, \ch{Fe2O3}, and \ch{Fe3O4}, i.e., \SI{1811}{\kelvin}, \SI{1838}{\kelvin}, and \SI{1870}{\kelvin}, respectively, are all greater than that of \ch{FeO}.}\\
 
 Khaikin~\cite{Khaikin1970Ignition} proposed a generic ignition model for reactive metal particles considering the obstructing effect of an oxide layer on the reaction rate, but did not apply it for quantitatively estimating the ignition temperature of iron particles. A quantitative model was developed by Bolobov and Podlevskikh~\cite{BolobovPodlevskikh2001} to explain the ignition characteristics of a low-carbon steel foil as experimentally probed in Ref.~\cite{Bolobov2001}. Despite some ambiguity in the kinetic parameters for iron oxidation, the ignition temperatures predicted by Bolobov and Podlevskikh’s model~\cite{BolobovPodlevskikh2001} are fairly close to the experimental results with a discrepancy less than \SI{50}{\kelvin}. The current study aims to develop a model to quantitatively capture the ignition characteristics of fine iron particles. To this end, a physics-based kinetic model of solid-phase iron oxidation---a parabolic rate law reflecting that the oxidation rate is controlled by the lattice diffusion of Fe cations through oxide layers---is considered in the current ignition model. The values of the kinetic parameters were calibrated against the experimental measurement of growth rates of iron-oxide scales over a temperature range from \SI{973}{\kelvin} to \SI{1523}{\kelvin} obtained by Pa\"{i}dassi~\cite{Paidassi1958a}. Using this model, the effects of initial particle size, initial oxide layer thickness, inert gas species, radiative heat loss, and collective heating in a particulate suspension on the ignition characteristics of iron particles are examined. The model prediction of ignition temperature is compared with the experimental results of Grosse and Conway~\cite{Grosse1958} and Bolobov~\cite{Bolobov2001}. The transition behavior between kinetic-controlled and external-diffusion-controlled combustion regimes and the ignition delay times of iron particles are further examined. The possible sources of error of this model are also discussed.
 
\section{\label{Sect2}Iron oxidation kinetic model}

\subsection{\label{Sect2_1}Physics underlying solid-phase iron oxidation}

\begin{figure}[h!]
\centerline{\includegraphics[width=0.8\textwidth]{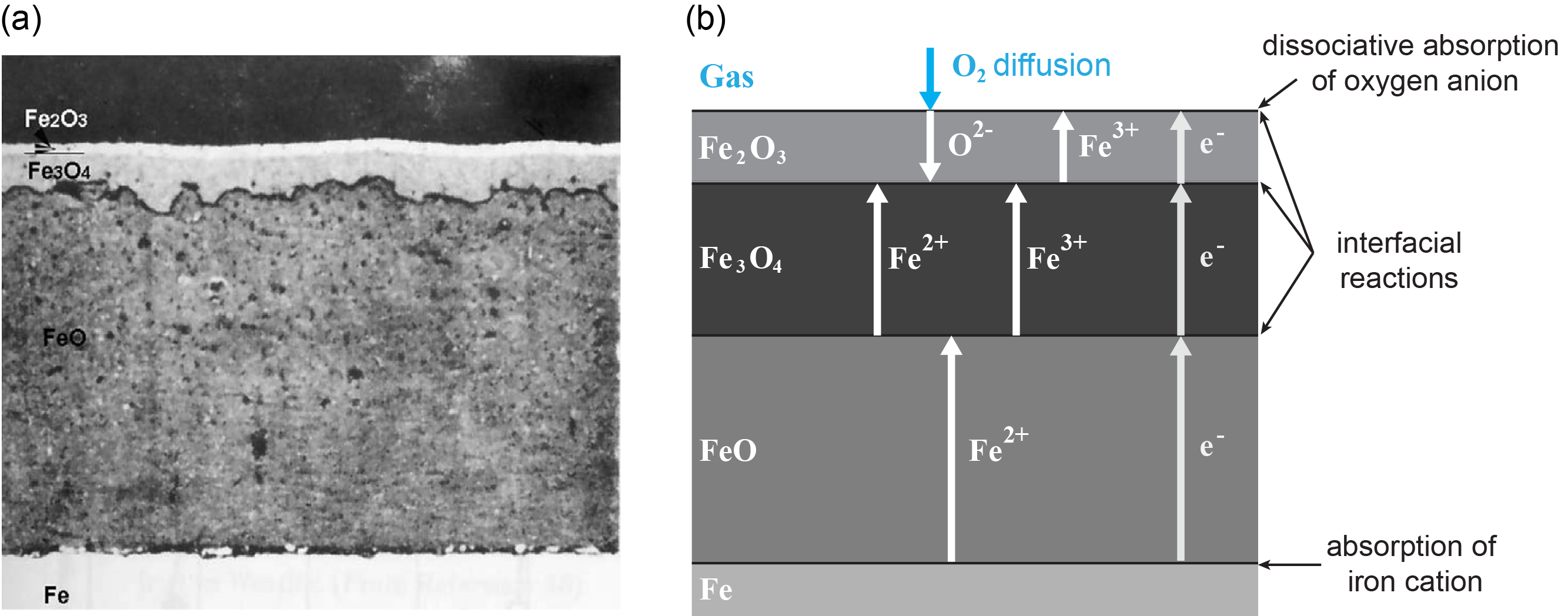}}
		\caption{(a) A sample cross-section view of a multilayered oxide scale grown on iron as shown by Pa\"{i}dassi \cite{Paidassi1958a}; (b) a schematic illustration showing the key processes underlying solid-phase iron oxidation.}
	\label{Fig1}
\end{figure}

Microscopic cross-section views of a solid-phase oxide scale grown on iron under isothermal conditions (over a range in temperature from $973~\mathrm{K}$ to $1523~\mathrm{K}$) in air were first obtained by Pa\"{i}dassi in the 1950s~\cite{Paidassi1958a}. Figure~\ref{Fig1}(a) is a sample image from Ref.~\cite{Paidassi1958a} showing that an iron-oxide scale consists of three compact layers of hematite (\ch{Fe2O3}), magnetite (\ch{Fe3O4}), and w\"{u}stite (\ch{FeO}) stacked from the gas-oxide interface to the oxide-iron interface. The relative thicknesses of the \ch{Fe2O3}, \ch{Fe3O4}, and \ch{FeO} layers with respect to the total thickness of the oxide scale are $1\%$, $4\%$, and $95\%$, respectively. As conceptually illustrated in Fig.~\ref{Fig1}(b), for a compact oxide film (i.e., without cracks or grain boundaries to facilitate oxygen transport to the iron core), solid-phase iron oxidation consists of the following processes:
\begin{enumerate}
  \item Diffusion of oxygen (\ch{O2}) molecules from the bulk gas to the gas-oxide interface.
  \item Incorporation of oxygen into the oxide scale (i.e., a dissociative absorption of \ch{O2} molecules as oxygen anions \ch{O^{2-}}).
  \item Absorption of iron into the oxide film as iron cations.
  \item Diffusion of ions and electrons through the oxide scale.
  \item Reactions at \ch{FeO}-\ch{Fe3O4}, \ch{Fe3O4}-\ch{Fe2O3}, and \ch{Fe2O3}-\ch{O2} interfaces.
\end{enumerate}
At relatively low temperatures, the diffusion of oxygen in the gas phase (Process~1) is much more rapid than the kinetic processes (2 to 5) in the solid phase. As the temperature of the particle increases, the solid-phase kinetics become increasingly rapid, and thus, the external diffusion of oxygen becomes the rate-limiting process. In this analysis, the oxidation rate is considered to be only controlled by the solid-phase kinetics for determining the critical conditions for particle ignition, i.e., the critical temperature required to trigger thermal runaway, in Sects.~\ref{Sect5_0}-\ref{Sect5_4}; a \textit{switch-type} reaction-rate model, considering the transition from kinetic- to external-diffusion-controlled combustion, is used describe the ignition behavior of iron particles beyond the critical temperature for thermal runaway, in Sects.~\ref{Sect5_5} and~\ref{Sect5_7}.\\

For a sufficiently thick oxide scale, the growth rate of each layer is controlled by the diffusion of ions subjected to the equilibrium activities of \ch{Fe} and \ch{O} at the interfaces~\cite{ChenYuen2003}. Considering the fact that the transport of electrons and the establishment of local equilibria at the interfaces are significantly more rapid than the diffusion of ions across the oxide layers, Wagner's theory~\cite{Wagner1933} relates the oxide growth rate to the diffusivity of ions. Since \ch{Fe} cations are significantly more mobile in \ch{FeO} and \ch{Fe3O4} layers than \ch{O} anions due to a large size of \ch{O} anion ($1.4~\mathrm{\AA}$)~\cite{Li2011EnergyEnviron}, the growth of oxide film is predominantly governed by the outward diffusion of \ch{Fe} cations~\cite{Himmel1953JOM,SmeltzerYoung1975}. The growth of the \ch{FeO} layer is the most rapid due to the fact that the diffusion coefficient of \ch{Fe} cations in \ch{FeO} is greater than that in \ch{Fe3O4} or \ch{Fe2O3}. Such a diffusion-controlled growth follows a parabolic rate law,
\begin{equation}
    \frac{\mathrm{d}X_i}{\mathrm{d}t} = \frac{k_{\mathrm{p},i}}{X_i}
    \label{Eq1}
\end{equation}
where $X_i$ is the thickness of an oxide layer, $k_{\mathrm{p},i}$ is the parabolic rate constant, and $i$ is the index of each oxide layer. Yurek~\textit{et al}.~\cite{Yurek1974} and Garnaud and Rapp~\cite{GarnaudRapp1977} extended Wagner's theory to show that the growth of a multilayered oxide film on iron follows a parabolic rate law governed by the diffusion coefficients of \ch{Fe} in \ch{FeO} and \ch{Fe3O4}. It is of importance to note that, as revealed by Goursat and Smeltzer~\cite{Goursat1973kinetics}, the oxidation kinetics of \ch{FeO} is independent of ambient oxygen concentration over a temperature range (at least) from $1073~\mathrm{K}$ to $1273~\mathrm{K}$ for a partial pressure of \ch{O2} ($p_\mathrm{\ch{O2}}$) above $4 \times 10^{-4}~\mathrm{atm}$. As the activities of \ch{Fe} and \ch{O} at the \ch{Fe}-\ch{FeO}, \ch{FeO}-\ch{Fe3O4}, and \ch{Fe3O4}-\ch{Fe2O3} interfaces are fixed by the phase equilibria, the growth of \ch{FeO} and \ch{Fe3O4} are barely affected by the ambient \ch{O2} concentration~\cite{Young2008}. In this analysis, the kinetic rates of \ch{FeO} and \ch{Fe3O4} formation are assumed to be independent of ambient \ch{O2} concentration as long as $p_\mathrm{\ch{O2}}$ is greater than zero. The implication of this assumption on the ignition process---the transition from a kinetic-controlled combustion to an \ch{O2}-diffusion-controlled (also referred to as external-diffusion-controlled) combustion---of an iron particle is discussed in Sect.~\ref{Sect5_5}.\\

Since the growth of the \ch{Fe2O3} layer is significantly slower in comparison to the growth of the \ch{FeO} and \ch{Fe3O4} layers, the energy release due to the formation \ch{Fe2O3} is negligible. The formation of this thin \ch{Fe2O3} layer is thus neglected in the current analysis. It is, however, of importance to understand the role played by the hematite layer on the formation of the inner oxidized layers. There is experimental evidence~\cite{Kofstad1972,Atkinson1985} indicating that the formation of \ch{Fe2O3} is due to both an outward diffusion of \ch{Fe} cations and an inward diffusion of \ch{O} anions at comparable rates (as illustrated in Fig.~\ref{Fig1}(b)), possibly along cracks in the hematite layer. Goursat and Smeltzer~\cite{Goursat1973kinetics} showed that the \ch{Fe2O3} layer formed upon the \ch{Fe3O4} layer is a dense ``forest'' of whiskers and platelets. The growth rate of this irregular hematite layer adapts to the ambient \ch{O2} concentration, establishing an equilibrium at the \ch{Fe3O4}-\ch{Fe3O2} interface, and thus, making the growth rates of \ch{FeO} and \ch{Fe3O4} independent of \ch{O2} concentration.\\

\subsection{\label{Sect2_2}Calibration of the kinetic model}

\begin{figure}[h!]
\centerline{\includegraphics[width=0.6\textwidth]{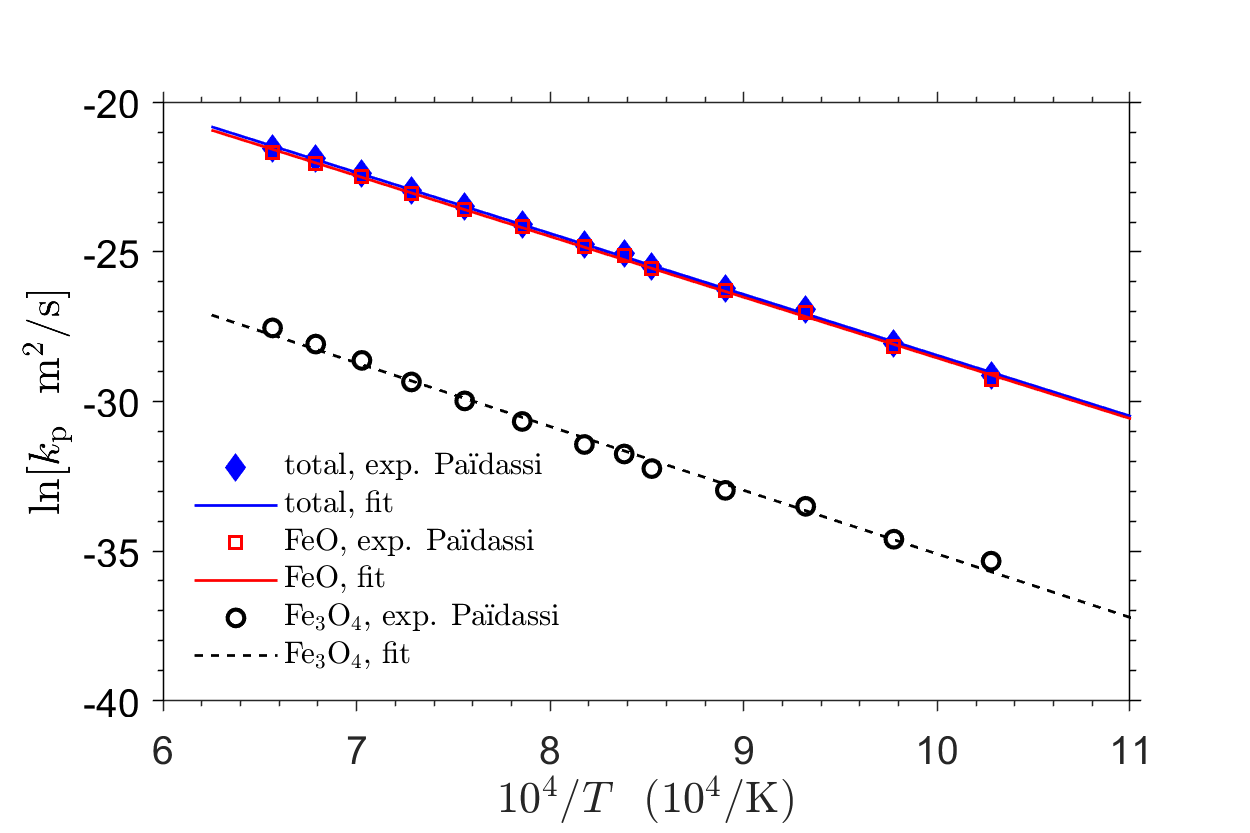}}
		\caption{The natural logarithmic values of the experimental data of $k_{\mathrm{p},i}$ for the growth of total oxide scale thickness (blue diamond), \ch{FeO} layer thickness (red square), and \ch{Fe3O4} layer thickness (black circle)  plotted as a function of the reciprocal of temperature, $1/T$, and a fitting line showing a linear regression of the corresponding data. Note that the \ch{FeO} and total oxide growth rates are effectively equal due to the much slower growth rate of \ch{Fe3O4}.}
	\label{Fig2}
\end{figure}

The experimental data of Pa\"{i}dassi~\cite{Paidassi1958a} reveal that, at a steady temperature, the thickness of each oxide layer $X_i$ increases with square root of time $\sqrt{t}$ at a constant rate, indeed following a parabolic rate law. The parabolic rate constant $k_i$ can be measured as the slope of the fitted line to the experimental data of $X_i$ as a function of $\sqrt{t}$ at each temperature, i.e., $X_i = k_i \sqrt{t} + X_{i,0}$, as reported in Table~1 of Ref.~\cite{Paidassi1958a}. Note that $k_i$ can be converted to the rate constant $k_{\mathrm{p},i}$ in Eq.~\ref{Eq1} via the relation as follows:
\begin{equation}
    k_{\mathrm{p},i} = \frac{{k_i}^2}{2}
    \label{Eq2}
\end{equation}
The natural logarithmic values of the experimental data of $k_{\mathrm{p},i}$ for the growth of total oxide film thickness (blue diamond), \ch{FeO} layer thickness (red square), and \ch{Fe3O4} layer thickness (black circle) are plotted in Fig.~\ref{Fig2} as a function of the reciprocal of temperature, $1/T$. These data points demonstrate a linear regression, i.e., the parabolic rate constant increases exponentially with temperature, indicating that the diffusion of \ch{Fe} cations in the oxide layers is an activated process. The temperature dependence of  $k_{\mathrm{p},i}$ can be described by an Arrhenius function as follows,
\begin{equation}
    k_{\mathrm{p},i}(T) = k_{0,i} \mathrm{Exp} \left( \frac{-T_{\mathrm{a},i}}{T} \right)
    \label{Eq3}
\end{equation}
wherein the values of pre-exponential factor $k_{0,i}$ and activation temperature $T_{\mathrm{a},i}$ for the growth of \ch{FeO} and \ch{Fe3O4} layers can be calibrated via line fitting to the plotted data points as reported in Table~\ref{Tab1}. 

\begin{table}
\begin{center}
\caption{Calibrated parameters for the parabolic growth rate of \ch{FeO} and \ch{Fe3O4}}
\label{Tab1}
\begin{tabular}{| c | c | c |}
\hline
- & $k_0~(\SI{}{\square\metre\per\second})$ & $T_\mathrm{a}~(\SI{}{\kelvin})$\\
\hline
\ch{FeO} & $2.670 \times 10^{-4}$ & $20319$\\
\ch{Fe3O4} & $1.027 \times 10^{-6}$ & $21310$\\
\hline
\end{tabular}
\end{center} 
\end{table}

\section{\label{Sect3}Model of iron particle ignition}

In the current analysis, a thermophysical model based on the mass and energy balance equations with an empirically calibrated kinetic model of iron oxidation is used to describe the ignition process of an isolated iron particle and a suspension of iron particles. The detailed formulation and major assumptions made in this model are presented this section. 

\begin{figure}[h!]
\centerline{\includegraphics[width=0.8\textwidth]{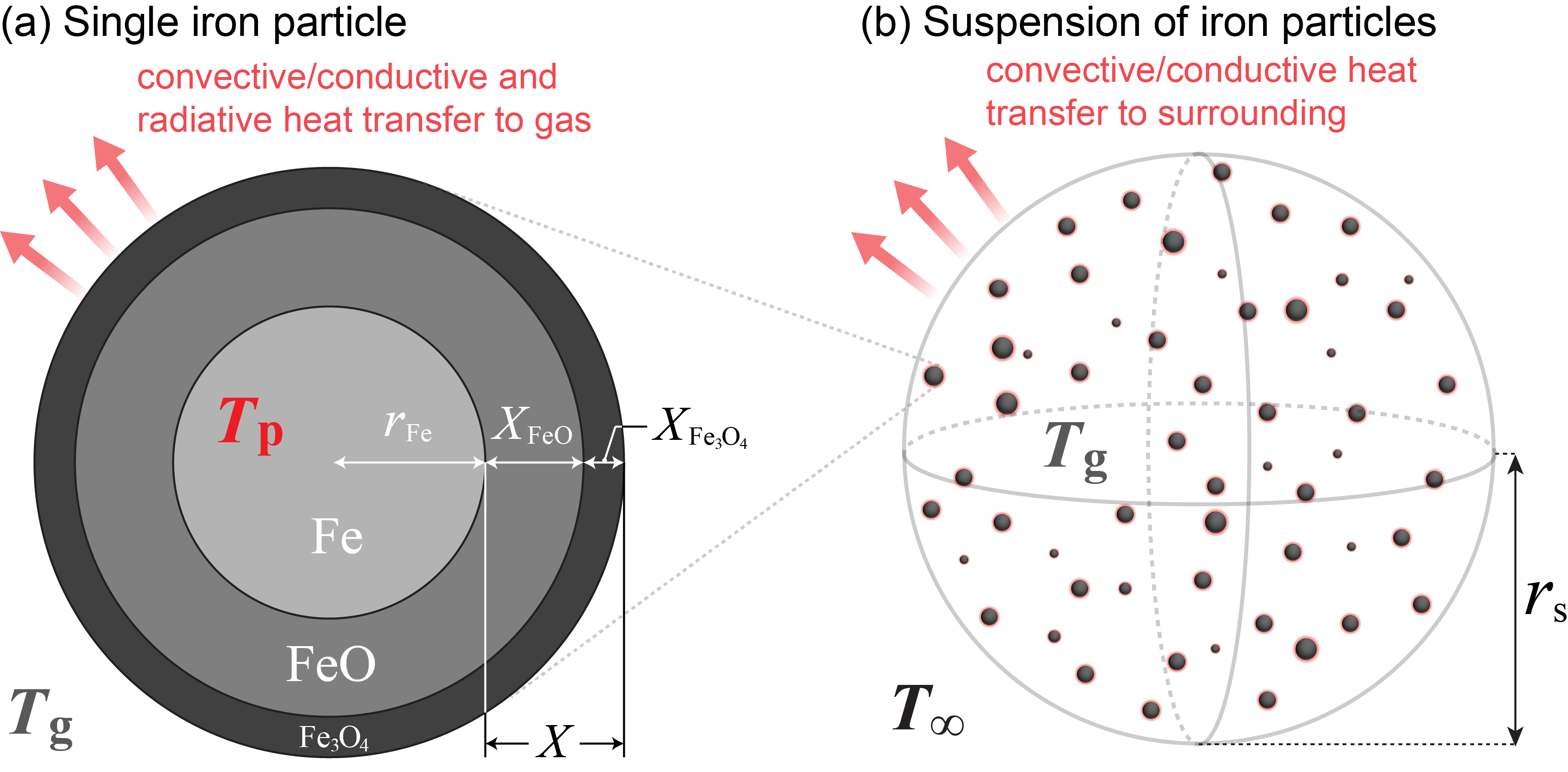}}
		\caption{Schematic of the thermophysical model of (a) an isolated particle consisting of a core of pure \ch{Fe}, an inner oxide layer of \ch{FeO}, and an outer oxide layer of \ch{Fe3O4} and (b) a spherical suspension of reacting iron particles.}
	\label{Fig3}
\end{figure}

\subsection{\label{Sect3_1}Model formulation for an isolated iron particle}

This model considers a spherical particle consisting of a core of iron, an inner \ch{FeO} layer, and an outer \ch{Fe3O4} layer as illustrated in Fig.~\ref{Fig3}(a). For the temperature range over which the oxidation kinetic model is calibrated, i.e., $973~\mathrm{K}$ to $1523~\mathrm{K}$, the thickness of \ch{Fe2O3} is about $1~\%$ of the total thickness of the oxide film. The formation of \ch{Fe2O3} is thus neglected in this model. Since the thermal conductivities of \ch{Fe} and \ch{FeO} are greater than that of the surrounding gas by two to three orders of magnitude, a typical Biot number of a fine iron particle is on the order of $0.001$-$0.01$. Hence, a uniform particle temperature, $T_\mathrm{p}$, can be assumed. The particle exchanges heat with the ambient gas at a temperature $T_\mathrm{g}$ via convective and radiative heat transfer. Other major assumptions on which the current model is based are summarized in the list below:
\begin{enumerate}
  \item The densities of solid iron, \ch{FeO}, \ch{Fe3O4} are considered as constant values, i.e., the effect of thermal expansion is neglected. Given the fact that, in the temperature range between \SI{900}{\kelvin} and \SI{1200}{\kelvin}, which is relevant to the current analysis, the coefficients of thermal expansion of \ch{Fe}, \ch{FeO}, and \ch{Fe3O4} vary over approximately the same range on the order of $1 \times 10^{-6} \mathrm{K}^{-1}$~\cite{Carter1959,Holcomb2019,Kozlovskii2019}, oxide layer cracking due the mismatch in thermal expansion between the oxide shell and the metal core is unlikely to core for iron particles.
  \item The phase changes from $\alpha$-\ch{Fe} to $\gamma$-\ch{Fe} and from $\gamma$-\ch{Fe} to $\delta$-\ch{Fe} are not considered since the associated latent heats are much less than the specific energy release of \ch{FeO} and \ch{Fe3O4} formation. The thermodynamic properties of iron used in the calculation are those of $\alpha$-\ch{Fe}.
  \item Curvature effect on the oxide layer growth is neglected as the oxide layer thickness is much less than the particle radius during the ignition process.
  \item The flow velocity of the ambient gas relative to the particle is negligible, i.e., the corresponding Reynolds number is zero.
\end{enumerate}

The mass balance equations for \ch{Fe}, \ch{FeO}, and \ch{Fe3O4} can be related to the parabolic kinetic model (Eqs.~\ref{Eq1} and~\ref{Eq3}) as follows,
\begin{equation}
    \frac{\mathrm{d} m_{\ch{Fe}}}{\mathrm{d} t} = - \nu_{\ch{Fe}/\ch{FeO}} \frac{\mathrm{d} m_{\ch{FeO}}}{\mathrm{d} t} - \nu_{\ch{Fe}/\ch{Fe3O4}} \frac{\mathrm{d} m_{\ch{Fe3O4}}}{\mathrm{d} t}
    \label{Eq4}
\end{equation}
\begin{equation}
    \frac{\mathrm{d} m_{\ch{FeO}}}{\mathrm{d} t} = \rho_{\ch{FeO}} A_{\ch{FeO}} \frac{\mathrm{d} X_{\ch{FeO}}}{\mathrm{d} t} = \frac{\rho_{\ch{FeO}} A_{\ch{FeO}} k_{0,\ch{FeO}}}{X_{\ch{FeO}}} \; \mathrm{Exp} \left( \frac{-T_\mathrm{a,FeO}}{T_\mathrm{p}} \right)
    \label{Eq5}
\end{equation}
\begin{equation}
    \frac{\mathrm{d} m_{\ch{Fe3O4}}}{\mathrm{d} t} = \rho_{\ch{Fe3O4}} A_\mathrm{p} \frac{\mathrm{d} X_{\ch{Fe3O4}}}{\mathrm{d} t} = \frac{\rho_{\ch{Fe3O4}} A_\mathrm{p} k_{0,\ch{Fe3O4}}}{X_{\ch{Fe3O4}}} \; \mathrm{Exp} \left( \frac{-T_\mathrm{a,\ch{Fe3O4}}}{T_\mathrm{p}} \right)
    \label{Eq5add}
\end{equation}
where $\nu_{\ch{Fe}/\ch{FeO}}$ and $\nu_{\ch{Fe}/\ch{Fe3O4}}$ are the stoichiometric mass ratios. Note that the formation of \ch{FeO} takes place at the outer surface of the \ch{FeO} layer (with an area $A_{\ch{FeO}}=4 \pi (r_{\ch{Fe}}+X_{\mathrm{FeO}})^2$) and the formation of \ch{Fe3O4} at the outer surface of the particle (with an area $A_\mathrm{p}=4 \pi (r_{\ch{Fe}}+X_{\mathrm{FeO}}+X_{\mathrm{Fe3O4}})^2$). The ratio between the total oxide layer thickness and particle radius, $\delta$, and the ratio between the \ch{Fe3O4} layer thickness and the total oxide layer thickness, $\delta_{\ch{Fe3O4}}$, are defined as follows:\\
\begin{equation}
    \delta = \frac{X}{r_\mathrm{p}} = \frac{X_{\mathrm{FeO}} + X_{\ch{Fe3O4}}}{r_\mathrm{p}} \;\;\;\;\;\;\;\; \delta_{\ch{Fe3O4}} = \frac{X_{\ch{Fe3O4}}}{X}
    \label{Eq5add1}
\end{equation}
In this analysis, the initial value of $\delta_{\ch{Fe3O4}}$ is set to be $0.05$ as inferred from experimental measurements ~\cite{Paidassi1958a,Paidassi1958b}. Given the fact that the total thickness of the oxide layer is much thinner than the radius of the iron core, i.e., $X_\mathrm{0} \ll r_\mathrm{Fe,0}$ and $r_\mathrm{Fe,0} \approx r_\mathrm{p,0}$, the initial ratio between the oxide layer thickness and the particle radius, $\delta_0$, is approximately proportional to the initial mass fraction of the oxide layer via the follow relation:
\begin{equation}
    \frac{m_\mathrm{oxide,0}}{m_\mathrm{Fe,0}} \approx \frac{4 \pi {r_\mathrm{Fe,0}}^2 X_0 \rho_\mathrm{oxide}}{\frac{4}{3} \pi {r_\mathrm{Fe,0}}^3 \rho_\mathrm{Fe}}= \frac{3 \rho_\mathrm{oxide} X_0}{\rho_\mathrm{Fe} r_\mathrm{p,0}} = \frac{3 \rho_\mathrm{oxide} {\color{red}\delta_0}}{\rho_\mathrm{Fe}}.
\end{equation}\\
where $\rho_\mathrm{oxide}$ is the average density of the oxide layer. The initial mass fraction of iron oxides is an experimentally measurable quantity.\\

The total enthalpy (or internal energy) of the particle $H_\mathrm{p}$ is the sum of the enthalpy of each solid-phase species,
\begin{equation}
    H_\mathrm{p} = \sum_{i}^{N_\mathrm{s}} \frac{m_i h_i(T_\mathrm{p})}{W_i}
    \label{Eq6}
\end{equation}
where $N_\mathrm{s}$ is the number of solid-phase species, $i$ is the index of each species, and $W_i$ is the molar weight of the $i^\mathrm{th}$ species, and $h_i$ is the corresponding molar enthalpy as a function of temperature described by the Shomate equation based on the NIST Standard Reference Data~\cite{Chase_NIST}. Knowing the value of $H_\mathrm{p}$ and $m_i$, particle temperature $T_\mathrm{p}$ can be determined via Eq.~\ref{Eq6} following an iterative root-finding procedure. The energy balance equation of the particle tracks the rate of change in $H_\mathrm{p}$ as follows,
\begin{equation}
    \frac{\mathrm{d} H_\mathrm{p}}{\mathrm{d} t} = q_{\ch{FeO}} \frac{\mathrm{d} m_{\ch{FeO}}}{\mathrm{d} t} + q_{\ch{Fe3O4}} \frac{\mathrm{d} m_{\ch{Fe3O4}}}{\mathrm{d} t} - {\frac{h_{\ch{O2}}(T_\mathrm{g})}{W_{\ch{O2}}}} \frac{\mathrm{d} m_{\ch{O2}}}{\mathrm{d} t} - A_\mathrm{p} \left[ h_\mathrm{p} (T_\mathrm{p}-T_\mathrm{g}) + \sigma \epsilon ({T_\mathrm{p}}^4-{T_\mathrm{g}}^4) \right]
    \label{Eq7}
\end{equation}
The first two terms on the right-hand side of Eq.~\ref{Eq7} are the energy release rates due to the formation of \ch{FeO} and \ch{Fe3O4}, respectively, where $q_{\ch{FeO}}$ and $q_{\ch{Fe3O4}}$ are values of the specific energy release that are related to the enthalpy of formation $H^{\circ}_{\mathrm{f}}$ of the oxide products as $q_{\ch{FeO}} = -\Delta H^{\circ}_{\mathrm{f},\ch{FeO(S)}} / W_{\ch{FeO}}$ and $q_{\ch{Fe3O4}} = -\Delta H^{\circ}_{\mathrm{f},\ch{Fe3O4(S)}} / W_{\ch{Fe3O4}}$. The third term in Eq.~\ref{Eq7} accounts for the increase in the enthalpy of the particle due to the incorporation of the mass of gaseous oxygen\footnote{The negative value of the mass consumption rate of oxygen, $\mathrm{d} m_{\ch{O2}}/\mathrm{d} t$, makes the oxygen enthalpy term in the energy balance positive.}, wherein $h_{\ch{O2}}$ is the molar enthalpy of \ch{O2} given by the NASA Thermodynamics Database~\cite{Mcbride1993}. The mass consumption rate of \ch{O2} is calculated as follows,
\begin{equation}
    \frac{\mathrm{d} m_{\ch{O2}}}{\mathrm{d} t} = - \nu_{\ch{O2}/\ch{FeO}} \frac{\mathrm{d} m_{\ch{FeO}}}{\mathrm{d} t} - \nu_{\ch{O2}/\ch{Fe3O4}} \frac{\mathrm{d} m_{\ch{Fe3O4}}}{\mathrm{d} t}
    \label{Eq8}
\end{equation}
where $\nu_{\ch{O2}/\ch{FeO}}$ and $\nu_{\ch{O2}/\ch{Fe3O4}}$ are the stoichiometric mass ratios.
\\

The fourth and fifth terms in Eq.~\ref{Eq7} are the rates of convective and radiative heat transfer between the particle and the ambient gas, respectively. The coefficient of convective heat transfer of the particle $h_\mathrm{p}$ can be determined as follows,
\begin{equation}
    h_\mathrm{p} = \frac{\mathrm{Nu} \lambda_\mathrm{g}}{d_\mathrm{p}}
    \label{Eq9}
\end{equation}
where $\mathrm{Nu}$ and $d_\mathrm{p}$ are the Nusselt number and diameter of the particle, respectively, and $\lambda_\mathrm{g}$ is the thermal conductivity of the ambient gas mixture of \ch{O2} and an inert species (i.e., \ch{N2},  \ch{Ar}, \ch{He}, and \ch{Xe} considered in this analysis). The thermal conductivity of each gaseous species $\lambda_i$ as a function of temperature can be determined using the NASA polynomials~\cite{Mcbride1993}. The transport properties of the gaseous species are evaluated in the particle-gas boundary layer, $T_\mathrm{sf}$. A ``two-third law''~\cite{Hubbard1975,Hazenberg2021} is used to estimate $T_\mathrm{sf}$, i.e., $T_\mathrm{sf} = (2 T_\mathrm{p} + T_\mathrm{g}) / 3$, in this paper.\footnote{Due to the fact that the difference between $T_\mathrm{p}$ and $T_\mathrm{g}$ is small during the ignition process, different ways of estimating $T_\mathrm{sf}$ does not make a significant difference in the prediction of the model.} The mixture-averaged thermal conductivity $\lambda_\mathrm{g}$ can then be calculated as follows,
\begin{equation}
    \lambda_\mathrm{g} = \frac{1}{2} \left( \sum_i^{N_\mathrm{g}} X_i \lambda_i + \frac{1}{\sum_i^{N_\mathrm{g}} X_i / \lambda_i} \right)
    \label{Eq10}
\end{equation}
where $N_\mathrm{g}$ is the number of species in the gaseous mixture and $X_i$ is the mole fraction of the $i^\mathrm{th}$ species. The Nusselt number can be estimated using the Fr\"{o}ssling (or Ranz-Marshal) correlation:
\begin{equation}
    \mathrm{Nu}=2+0.552 \mathrm{Re}^{\frac{1}{2}} \mathrm{Pr}^{\frac{1}{3}}.
    \label{Eq11new}
\end{equation}
As a result of Assumption~4, the Nusselt number equals to two (i.e., $\mathrm{Nu}=2$) for a spherical object in a quiescent medium, and thus, $h_\mathrm{p} = \lambda_\mathrm{g} / r_\mathrm{p}$. Note that, although the term $h_\mathrm{p}(T_\mathrm{p}-T_\mathrm{g})$ describes the rate of convective heat transfer, it is referred to as ``conductive heat loss'' when discussing the results of the current analysis given the assumption of a quiescent gas medium. In the radiative heat transfer term, $\sigma$ and $\epsilon$ denote the Stefan-Boltzmann constant and the emissivity of the oxide layer (\ch{Fe3O4}) covering the particle surface, respectively. A constant value of $\epsilon=0.88$ is considered in this analysis as the total emissivity of magnetite varies slightly from $0.85$ to $0.89$ as temperature increases from \SI{773}{\kelvin} to \SI{1473}{\kelvin}~\cite{Burgess1915}. Note that, since the total emissivity of hematite varies from $0.75$ to $0.85$ over a range in temperature from \SI{850}{\kelvin} to \SI{1300}{\kelvin}~\cite{Touloukian1972,Jones2019}, the current model may slightly overestimates the radiative heat transfer rate of an iron particle covered by a thin layer of \ch{Fe2O3}. The values of the key properties of the solid species are provided in Table~\ref{Tab2}.

\begin{table}
\begin{center}
\caption{Properties of \ch{Fe}, \ch{FeO}, and \ch{Fe3O4}}
\label{Tab2}
\begin{tabular}{| c | c | c | c |}
\hline
Property & Value & Unit\\
\hline
$\rho_{\ch{Fe}}$ & 7874 & \SI{}{\kilo\gram\per\cubic\meter}\\
\hline
$\rho_{\ch{FeO}}$ & 5745 & \SI{}{\kilo\gram\per\cubic\meter}\\
\hline
$\rho_{\ch{Fe3O4}}$ & 5170 & \SI{}{\kilo\gram\per\cubic\meter}\\
\hline
$q_{\ch{FeO}}$ & 3.787 & \SI{}{\mega\joule\per\kilo\gram}\\
\hline
$q_{\ch{Fe3O4}}$ & 4.841 & \SI{}{\mega\joule\per\kilo\gram}\\
\hline
$\epsilon$ & 0.88 & -\\
\hline
\end{tabular}
\end{center} 
\end{table}

\subsection{\label{Sect3_2}Model formulation for a suspension of iron particles}

This model considers a spherical cloud of iron particles suspended in a gas mixture of \ch{O2} and an inert species at constant pressure. The radius of this spherical suspension (denoted as $r_\mathrm{s}$ shown in Fig.~\ref{Fig3}) will increase as a result of thermal expansion of the gas mixture. The model monitors the mass and heat transfer between the particles and the gas mixture (as described by the model for an isolated particle in Sect.~\ref{Sect3_1}) and the heat transfer between the gas mixture and the surrounding gas at a temperature $T_{\infty}$ outside the suspension. Several simplifying assumptions are made in order to better focus this study on the key parameters controlling the ignition characteristics of an iron-particle suspension:
\begin{enumerate}
  \item A monodisperse suspension, i.e., all of the particles are of the same initial size.
  \item The total volume occupied by the particles is negligible compared to the total volume of the suspension.
  \item Radiative heat transfer is neglected.
  \item The suspension expands freely at a constant pressure of \SI{1}{atm}.
  \item No slip between the expanding gas mixture and particles, i.e., the spatial distribution of particles expands with the gas mixture so that the volume of the entire suspension increases due to thermal expansion of the gas.
  \item Flow velocity of the surrounding gas relative to the suspension is zero.
\end{enumerate}
Knowing the initial radius of each iron particle, $r_{\ch{Fe},0}$, the initial, total mass of iron particles in a suspension, $m_\mathrm{\ch{Fe},tot,0}$, can be related to the number of particles $N_\mathrm{p}$ as follows:
\begin{equation}
    m_\mathrm{\ch{Fe},tot,0} =  \frac{4 \pi N_\mathrm{p} \rho_{\ch{Fe}} {r_{\ch{Fe},0}}^3}{3}.
    \label{Eq11}
\end{equation}
Although the overall stoichiometry of an iron-oxygen mixture is not uniquely defined, due to multiple possible oxidation states, an equivalence ratio considering \ch{FeO} as the only product can be defined as follows:
\begin{equation}
    \phi_{\ch{FeO}} = \frac{m_\mathrm{\ch{Fe},tot,0} / m_\mathrm{\ch{O2},0}}{2 W_{\ch{Fe}} / W_{\ch{O2}}}.
    \label{Eq12}
\end{equation}
The initial mass of \ch{O2} in the suspension $m_\mathrm{\ch{O2},0}$ is related to the initial temperature, $T_{\mathrm{g},0}$, pressure, $p_0$, and \ch{O2} mass fraction $Y_{\ch{O2},0}$:
\begin{equation}
    m_\mathrm{\ch{O2},0} = V_{\mathrm{s,0}} \rho_{\mathrm{g,0}} Y_{\ch{O2},0} = \frac{p_0 V_{\mathrm{s,0}} Y_{\ch{O2},0} }{\mathcal{R}_\mathrm{g} T_{\mathrm{g},0}},
    \label{Eq13}
\end{equation}
where $\mathcal{R}_\mathrm{g}$ is the specific gas constant of the gas mixture, $\rho_{\mathrm{g,0}}$ is the initial gas density, and $V_{\mathrm{s,0}}$ is the initial suspension volume. Thus, by specifying $T_{\mathrm{g},0}$, $p_0$, $Y_{\ch{O2},0}$, $r_{\ch{Fe},0}$, $r_{\mathrm{s},0}$, and $\phi_{\ch{FeO}}$, one can calculate $N_\mathrm{p}$, $m_\mathrm{\ch{Fe},tot,0}$, and $m_\mathrm{\ch{O2},0}$ via Eqs.~\ref{Eq11}-\ref{Eq13} as input parameters to this suspension model.\\

Based on the above-listed assumptions, the gas-phase energy balance equation of a suspension is formulated as follows:
\begin{equation}
    \frac{\mathrm{d} H_\mathrm{g}}{\mathrm{d} t} = N_\mathrm{p} A_\mathrm{p} h_\mathrm{p} (T_\mathrm{p}-T_\mathrm{g}) + A_\mathrm{s} h_\mathrm{s} (T_\mathrm{\infty}-T_\mathrm{g}).    
\end{equation}
The total enthalpy of the gas mixture $H_\mathrm{g}$ is the sum of the enthalpies of \ch{O2} and an inert species is,
\begin{equation}
    H_\mathrm{g} = \frac{m_{\ch{O2}} h_{\ch{O2}}(T_\mathrm{g})}{W_{\ch{O2}}} + \frac{m_\mathrm{I} h_\mathrm{I}(T_\mathrm{g})}{W_\mathrm{I}},
\end{equation}
where the subscript ``I'' denotes the inert species (i.e., \ch{N2}, \ch{Ar}, \ch{He}, and \ch{Xe} considered in this study). The mass balance equation of \ch{O2} in the suspension is formulated as:
\begin{equation}
    \frac{\mathrm{d} m_{\ch{O2}}}{\mathrm{d} t} = - N_\mathrm{p} (\nu_{\ch{O2}/\ch{FeO}} \frac{\mathrm{d} m_{\ch{FeO}}}{\mathrm{d} t} + \nu_{\ch{O2}/\ch{Fe3O4}} \frac{\mathrm{d} m_{\ch{Fe3O4}}}{\mathrm{d} t}).
\end{equation}
As the gas temperature, $T_\mathrm{g}$, and the mass of \ch{O2} change over time, the volume of the suspension changes according to the ideal gas law. Note that an isobaric condition at \SI{1}{atm} is considered in all of the calculations with the suspension model.

\section{\label{Sect4}Sample results}

\subsection{\label{Sect4_1}An isolated particle}
\begin{figure}[h!]
\centerline{\includegraphics[width=0.55\textwidth]{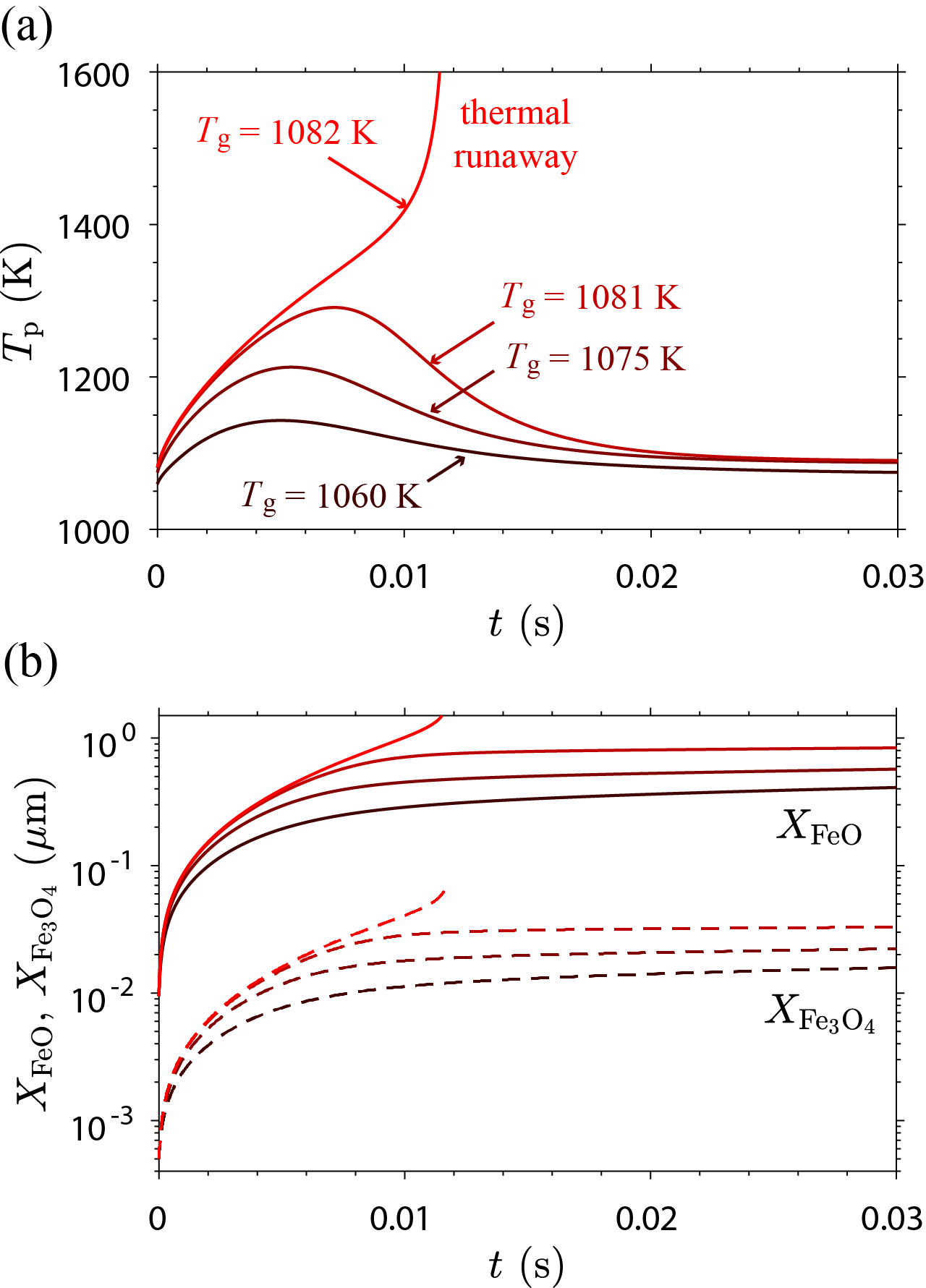}}
		\caption{Sample results showing the time histories of (a) particle temperature $T_\mathrm{p}$ and (b) the thicknesses of \ch{FeO} layer (solid curves) and \ch{Fe3O4} layer (dashed curves) at four different gas temperatures (i.e., from bottom to top, $T_\mathrm{g} = \SI{1060}{K}$, $\SI{1075}{K}$, $\SI{1081}{K}$, and $\SI{1082}{K}$) for an initial particle size $d_\mathrm{p,0}=\SI{20}{\micro\metre}$ with an initial oxide layer thickness of $X_0 = X_\mathrm{\ch{FeO},0}+X_\mathrm{\ch{Fe3O4},0}=\SI{10}{\nano\metre}$ in air ($X_{\ch{O2}}$:$X_{\ch{N2}}$=$0.21$:$0.79$).}
	\label{Fig4}
\end{figure}

Sample results of the calculations considering an isolated iron particle with an initial diameter $d_\mathrm{p,0}=\SI{20}{\micro\metre}$ and an initial oxide layer thickness $X_0 = X_\mathrm{\ch{FeO},0}+X_\mathrm{\ch{Fe3O4},0}=\SI{10}{\nano\metre}$ ($\delta_0=10^{-3}$) in air ($X_{\ch{O2}}$:$X_{\ch{N2}}$=$0.21$:$0.79$) are plotted in Fig.~\ref{Fig4}. The subfigures (a) and (b) show the time histories of particle temperature $T_\mathrm{p}$ and the thicknesses of the \ch{FeO} layer (solid curves) and \ch{Fe3O4} layer (dashed curves), respectively. Each curve corresponds to a case with a different constant gas temperature $T_\mathrm{g}$. Particle temperature is initialized to be equal to $T_\mathrm{g}$ for all of the cases. The initial thickness of the \ch{FeO} layer is $95\%$ of $X_0$. For the cases with $T_\mathrm{g} \leq \SI{1081}{\kelvin}$, as shown in Fig.~\ref{Fig4}(a), $T_\mathrm{p}$ non-monotonically varies over time, exhibiting an increase to a peak value followed by an asymptotic decrease to $T_\mathrm{g}$. Figure.~\ref{Fig4}(b) shows that the oxide layer thicknesses for $T_\mathrm{g} \leq \SI{1081}{\kelvin}$ reach a quasi-plateau after an initial increase until approximately \SI{0.01}{\second}. For the case with $T_\mathrm{g}=\SI{1082}{\kelvin}$, $T_\mathrm{p}$ increases monotonically with time as shown in Fig.~\ref{Fig4}(a). This increase in temperature becomes abrupt after approximately \SI{0.01}{\second}. The calculation was stopped once $T_\mathrm{p}$ reaches the melting point of \ch{FeO} (i.e., \SI{1650}{\kelvin}). At this stopping point, the slopes of the $X_{\ch{FeO}}$ and $X_{\ch{Fe3O4}}$ curves (shown in Fig.~\ref{Fig4}(b)) indicate that oxide layer thicknesses continue to increase instead of asymptotically approaching a plateau value.

\subsection{\label{Sect4_2}Suspension of particles}
\begin{figure}[h!]
\centerline{\includegraphics[width=0.55\textwidth]{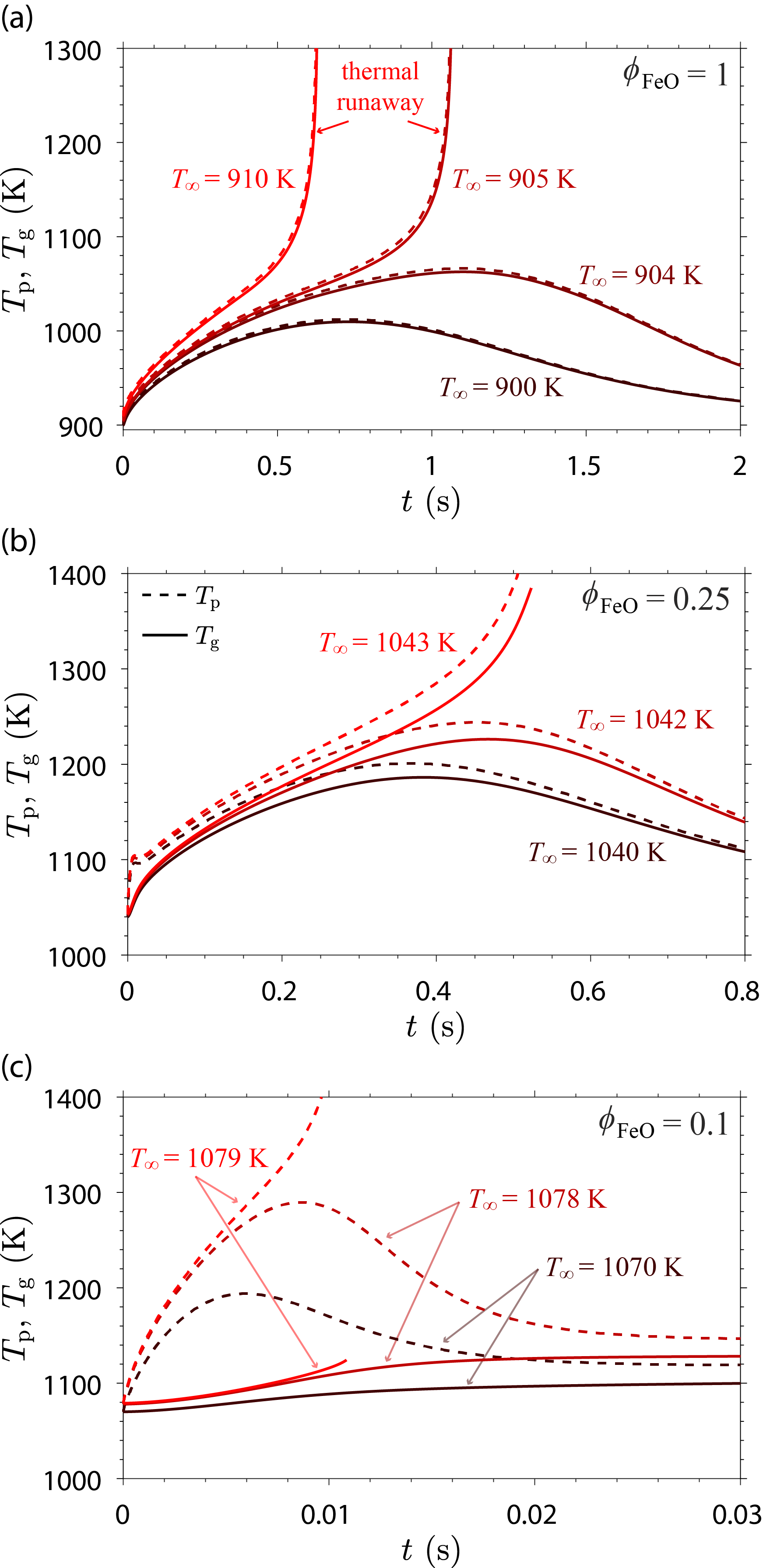}}
		\caption{Sample results showing the time histories of gas temperature $T_\mathrm{g}$ (solid curves) and particle temperature $T_\mathrm{p}$ (dashed curves) of a suspension of iron particles with an initial suspension radius $r_\mathrm{s,0}=\SI{1}{\centi\meter}$, particle size $d_\mathrm{p,0}=\SI{20}{\micro\metre}$, and oxide layer thickness of $X_0 =\SI{10}{\nano\metre}$ in air ($X_{\ch{O2}}$:$X_{\ch{N2}}$=$0.21$:$0.79$) at equivalence ratios (a) $\phi_{\mathrm{FeO}}=1$, (b) $\phi_{\mathrm{FeO}}=0.25$, and (c) $\phi_{\mathrm{FeO}}=0.1$ subjected to different surrounding gas temperatures.}
	\label{Fig5}
\end{figure}

Sample suspension model results shown in Fig.~\ref{Fig5} are for the cases with an initial suspension radius $r_\mathrm{s,0}=\SI{1}{\centi\meter}$, particle size $d_\mathrm{p,0}=\SI{20}{\micro\metre}$, and oxide layer thickness of $X_0 =\SI{10}{\nano\metre}$ in air at equivalence ratios $\phi_{\mathrm{FeO}}=1$, $0.25$, and $0.1$. The particle and gas temperatures of a suspension are initialized to be equal to the surrounding gas temperature $T_{\infty}$, and are plotted as dashed and solid curves in Fig.~\ref{Fig5}, respectively. Each pair of $T_\mathrm{p}$ and $T_\mathrm{g}$ curves correspond to a case with a constant $T_{\infty}$. The curve of $T_\mathrm{p}$ is slightly above that of $T_\mathrm{g}$ for all of the cases. For the cases with $\phi_{\mathrm{FeO}}=1$ as shown in Fig.~\ref{Fig5}(a), $T_{\infty} \leq \SI{904}{\kelvin}$, the resulting temperatures vary non-monotonically over time; for the cases with $T_{\infty} \geq \SI{905}{\kelvin}$, $T_\mathrm{p}$ and $T_\mathrm{g}$ increase abruptly after a gradual increase over early times. For the cases with an increasingly smaller $\phi_{\mathrm{FeO}}$, the separation between $T_\mathrm{p}$ and $T_\mathrm{g}$ increases.  For the cases with $\phi_{\mathrm{FeO}}=0.1$ as shown in Fig.~\ref{Fig5}(c), $T_\mathrm{p}$ increases much more rapidly than $T_\mathrm{g}$ does. In the case with $T_{\infty} = \SI{1079}{\kelvin}$, while $T_\mathrm{p}$ promptly increases, $T_\mathrm{g}$ slightly increases from the surrounding temperature.

\section{\label{Sect5}Analysis and discussion}

\subsection{\label{Sect5_0}Critical condition of thermal runaway}
A gradual increase in temperature followed by a rapid upsurge (as identified in some cases shown in Sect.~\ref{Sect4}) indicates a particle or suspension undergoing a thermal runaway. The bifurcation between the ``go'' and ``no-go'' cases is identified as a critical phenomenon of ignition. Based on the time histories resulting from the isolated particle and suspension models, critical values of $T_\mathrm{g}$ and $T_{\infty}$ above which an isolated particle or a suspension undergoes a thermal runaway, respectively, can be determined. This critical temperature for ignition is referred to as the \textit{ignition temperature} and denoted as $T_\mathrm{ign}$. In this study, $T_\mathrm{ign}$ is used as a metric to probe the effects of various factors contributing to the ignition process of iron particles. Note that the radiative heat transfer term in Eq.~\ref{Eq7} is neglected in most of the analysis shown in this section, with an exception of Sect.~\ref{Sect5_3} wherein the effect of radiation on iron particle ignition is examined.

\subsection{\label{Sect5_1}Effect of initial particle size and oxide layer thickness}

\begin{figure}[h!]
\centerline{\includegraphics[width=0.55\textwidth]{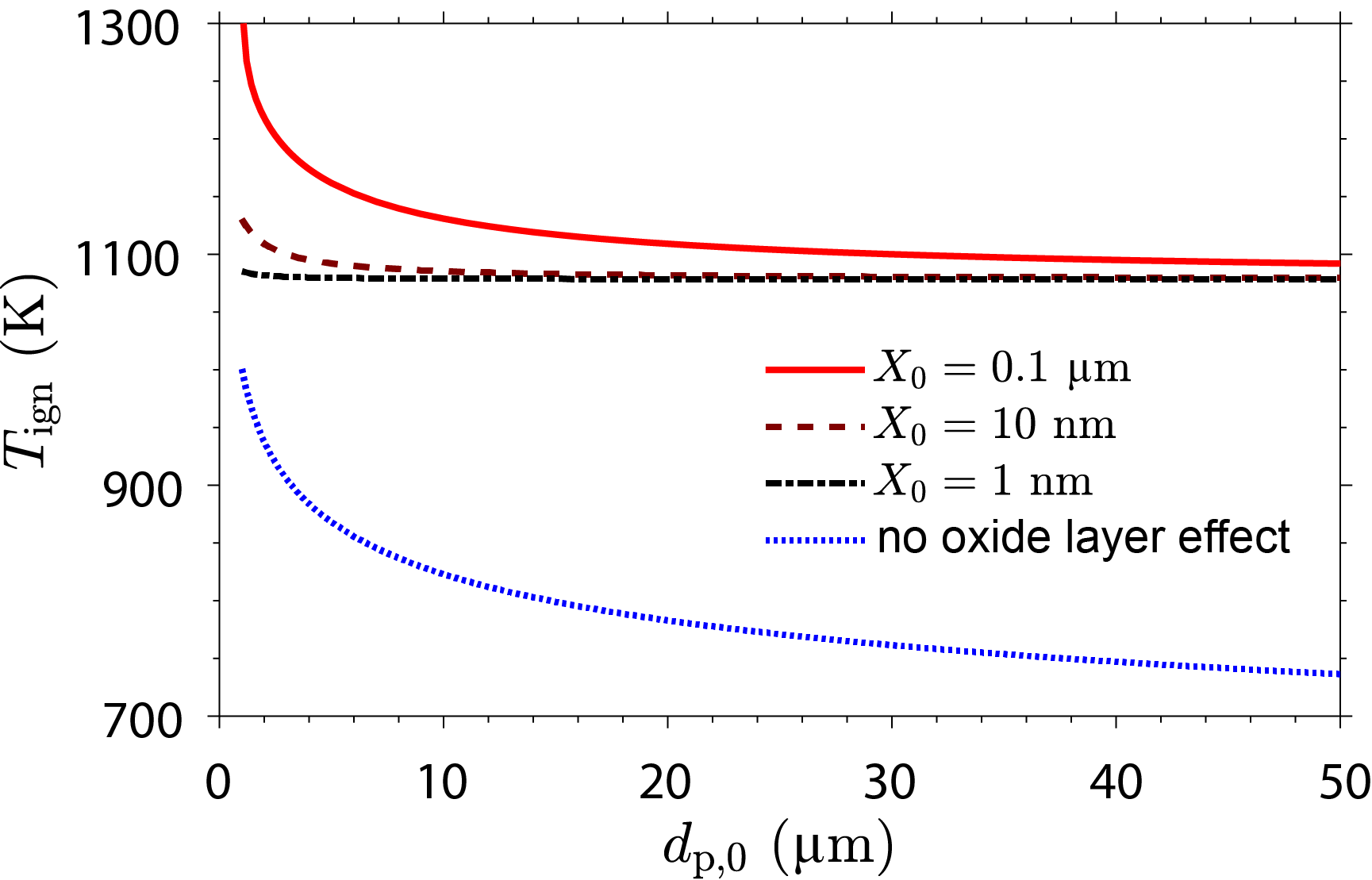}}
		\caption{Critical temperature for ignition, $T_\mathrm{ign}$, as a function of initial particle diameter, $d_\mathrm{p,0}$, with three different initial oxide layer thicknesses, i.e., $X_0=\SI{0.1}{\micro\meter}$ (solid red curve), $X_0=\SI{10}{\nano\meter}$ (dashed maroon curve), and $X_0=\SI{1}{\nano\meter}$ (dash-dotted black curve). Results from the current model for an isolated particle in air are compared to the result of a model~\cite{Hazenberg2021} neglecting the obstructing effect of oxide layer on the oxidation rate (plotted as the dotted blue curve).}
	\label{Fig6}
\end{figure}

The results of $T_\mathrm{ign}$ are plotted in Fig.~\ref{Fig6} as a function of initial particle diameter, $d_\mathrm{p,0}$, with three different initial oxide layer thicknesses for an isolated particle in air. The smallest particle size considered in this set of results is $d_\mathrm{p,0}=\SI{1}{\micro\meter}$. An ignition temperature significantly below the melting point of \ch{FeO} is found for all particle sizes above \SI{1}{\micro\meter} over a broad range of initial oxide layer thickness from \SI{1}{\nano\meter} to \SI{0.1}{\micro\meter}. This result suggests that micron-sized or larger iron particles, which were used in most of the experimental studies on iron-fuel flames~\cite{Tang2011PROCI,Wright2016,McRae2018,Palecka2019Perwaves,Toth2020,Palecka2020Perwaves,Poletaev2020}, very likely undergo an ignition process---transitioning from a kinetic-controlled reaction to an external-diffusion-controlled combustion---rather than a slow, kinetic-controlled burnout at the ambient gas temperature.\\

For the case with $X_0=\SI{0.1}{\micro\meter}$ (the solid red curve in Fig.~\ref{Fig6}), there is a significant decrease in $T_\mathrm{ign}$ as $d_\mathrm{p,0}$ increases from \SI{1}{\micro\meter} to approximately \SI{20}{\micro\meter}; $T_\mathrm{ign}$ nearly plateaus for further larger particles. Although qualitatively the same trend is identified for the case with a much thinner initial oxide layer of $X_0=\SI{10}{\nano\meter}$ (dashed maroon curve), $T_\mathrm{ign}$ decreases less significantly with an increase in $d_\mathrm{p,0}$ and reaches nearly the same plateau value ($\approx \SI{1082}{\kelvin}$) as that resulting from the case with $X_0=\SI{0.1}{\micro\meter}$. For an even thinner initial oxide layer of $X_0=\SI{1}{\nano\meter}$ (dash-dotted black curve), the entire curve of $T_\mathrm{ign}$ plateaus at \SI{1082}{\kelvin}, independent of initial particle size. This independence of $T_\mathrm{ign}$ on particle size (for sufficiently large particles) has been identified as a result of a parabolic oxidation rate law in Khaikin~\textit{et al}.'s~\cite{Khaikin1970Ignition} generic analysis of metal particle ignition, which is qualitatively explained as follows: For increasingly large particles, the increase in energy release rate (proportional to particle surface area, i.e., $\propto~{d_\mathrm{p,0}}^2$) surpasses the increase in the rate of conductive heat removal (proportional to particle size, i.e., $\propto~{d_\mathrm{p,0}}$). As a result, larger particles are more prone to ignition. If a parabolic oxidation rate law is considered, however, to heat up larger particles requires a greater amount of energy release associated with a thicker oxide layer grown, in return, impeding the oxidation rate. Hence, a large particle size has both enhancing and hindering effects on the ignition propensity of an iron particle. These mutually compensating effects result in the independence of $T_\mathrm{ign}$ on particle size for sufficiently large particles governed by a parabolic oxidation rate law. As a comparison, the result of a model (recently proposed by Hazenberg and Oijen~\cite{Hazenberg2021}) neglecting the impeding effect of an oxide layer on the oxidation rate is plotted as the dotted blue curve in Fig.~\ref{Fig6}. The resulting $T_\mathrm{ign}$ persistently decreases as $d_\mathrm{p,0}$ increases, manifesting only the enhancing effect of an increase in particle size. Therefore, prudence must be practiced when interpreting the flame simulation results using this likely oversimplified reaction model~\cite{Hazenberg2021}.\\

\begin{figure}[h!]
\centerline{\includegraphics[width=1.0\textwidth]{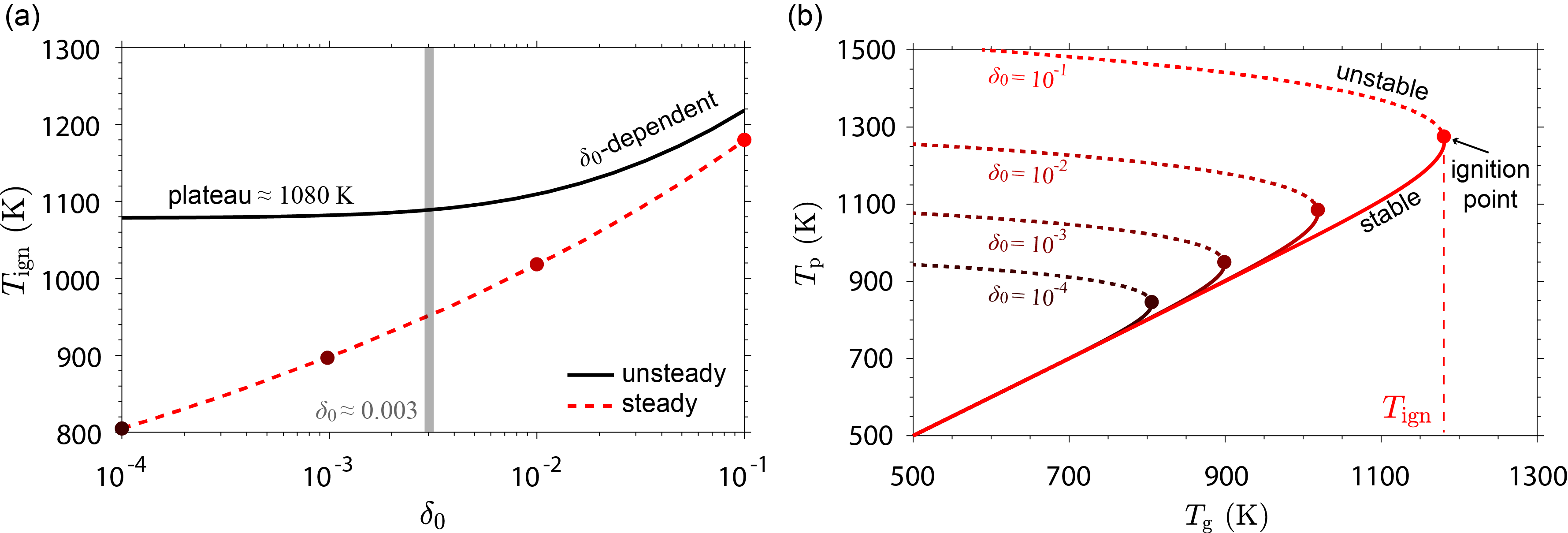}}
		\caption{(a) Ignition temperature, $T_\mathrm{ign}$, as a function of $\delta_0$, the ratio between initial oxide layer thickness and particle radius for an isolated iron particle in air. The results of the steady-state analysis based on the Semenov ignition criterion and the unsteady model are plotted as the dashed red curve and solid black curve, respectively. (b) The curves of $T_\mathrm{p}$ and $T_\mathrm{g}$ satisfying $\dot{q}_\mathrm{R}=\dot{q}_\mathrm{L}$ in the steady-state analysis with various $\delta_0$. The turning point of each curve marks the critical condition for ignition.}
	\label{Fig7}
\end{figure}

Further analysis is performed to understand the effect of initial oxide layer thickness. The criterion for thermal runway is that the rate of energy release from iron oxidation, $\dot{q}_\mathrm{R}$, must exceed the rate of heat loss to the surrounding gas, $\dot{q}_\mathrm{L}$, and is formulated as follows (neglecting radiative heat transfer):
\begin{equation}
\begin{split}
    \dot{q}_\mathrm{R} & > \dot{q}_\mathrm{L}\\
     q_{\ch{FeO}} \frac{\mathrm{d} m_{\ch{FeO}}}{\mathrm{d} t} + q_{\ch{Fe3O4}} \frac{\mathrm{d} m_{\ch{Fe3O4}}}{\mathrm{d} t}  & >  A_\mathrm{p} h_\mathrm{p} (T_\mathrm{p}-T_\mathrm{g}).
\end{split}
\label{Eq19}
\end{equation}
Note that the rate of energy increase due to the incorporation of gaseous oxygen (i.e., the third term on the right-hand side of Eq.~\ref{Eq7}) is also neglected in this simplified criterion. This simplification is based on the fact that the specific enthalpy associated with the incorporated \ch{O2} is insignificant compared to the specific energy release of \ch{FeO} and \ch{Fe3O4} formation. Considering the kinetic rate laws (Eqs.~\ref{Eq5} and \ref{Eq5add}) and the definition of $\delta$ and $\delta_{\ch{Fe3O4}}$ (Eq.~\ref{Eq5add1}), the thermal runaway criterion can be written as:
\begin{equation}
    \tilde{q}_{\ch{FeO}}  \; \mathrm{Exp} \left( \frac{-T_\mathrm{a,FeO}}{T_\mathrm{p}} \right) \frac{{(1-\delta \delta_{\ch{Fe3O4}})}^2 {\color{red}{r_\mathrm{p}}} }{\delta -\delta \delta_{\ch{Fe3O4}}} + \tilde{q}_{\ch{Fe3O4}}  \; \mathrm{Exp} \left( \frac{-T_\mathrm{a,\ch{Fe3O4}}}{T_\mathrm{p}} \right) \frac{{\color{red}{r_\mathrm{p}}}}{\delta \delta_{\ch{Fe3O4}}} > \lambda_\mathrm{g} (T_\mathrm{p} - T_\mathrm{g}) {\color{red}{r_\mathrm{p}}},
    \label{Eq20}
\end{equation}
where $\tilde{q}_{\ch{FeO}}=\rho_{\ch{FeO}} q_{\ch{FeO}} k_{0,\ch{FeO}}$ and $\tilde{q}_{\ch{Fe3O4}}=\rho_{\ch{Fe3O4}} q_{\ch{Fe3O4}} k_{0,\ch{Fe3O4}}$. Equation~\ref{Eq20} shows that the energy release rates (left-hand side) and conductive heat loss rate (right-hand side) both scale linearly with particle radius, $r_\mathrm{p}$. Thus, the thermal runaway criterion is independent of particle size when only conductive heat transfer is considered in the loss rate. As a result, $T_\mathrm{ign}$ exclusively depends on the initial oxide layer to particle radius ratio, $\delta_0$, for a given gas mixture (while assuming a constant value of $\delta_{\ch{Fe3O4},0}=0.05$) as shown in Fig.~\ref{Fig7}(a).\\

A simple steady-state analysis---neglecting the growth of the oxide layers over time---was first performed to estimate $T_\mathrm{ign}$ based on the Semenov ignition criterion (i.e., solving for $T_\mathrm{p}$ and $T_\mathrm{g}$ satisfying $\dot{q}_\mathrm{R}=\dot{q}_\mathrm{L}$ and $\mathrm{d}\dot{q}_\mathrm{R} / \mathrm{d} T_\mathrm{p}=\mathrm{d} \dot{q}_\mathrm{L} / \mathrm{d} T_\mathrm{p}$). The relation between $T_\mathrm{p}$ and $T_\mathrm{g}$ satisfying $\dot{q}_\mathrm{R}=\dot{q}_\mathrm{L}$ is plotted as the curves in Fig.~\ref{Fig7}(b) for various $\delta_0$. Each curve exhibits a turning-point behavior. The $T_\mathrm{p}$ and $T_\mathrm{g}$ at the turning point are found as the critical condition for ignition (i.e., satisfying $\mathrm{d}\dot{q}_\mathrm{R} / \mathrm{d} T_\mathrm{p}=\mathrm{d} \dot{q}_\mathrm{L} / \mathrm{d} T_\mathrm{p}$). The branches below and above the turning point correspond to stable and unstable solutions for $\dot{q}_\mathrm{R}=\dot{q}_\mathrm{L}$, respectively. As shown in Fig.~\ref{Fig7}(b), the turning point or $T_\mathrm{ign}$ increases with an increase in $\delta_0$.  This steady-state analysis prediction is also plotted in Fig.~\ref{Fig7}(a) as the dashed red curve. This trend is due to the fact that reaction rate is greater for an increasingly thin oxide layer, as described by the parabolic rate law. A noteworthy finding of the steady-state analysis is that, as shown in Fig.~\ref{Fig7}(b), the ignition point can be identified for all values of $\delta_0$ regardless of the particle size. In other words, there is no minimum size of an iron particle below which the particle cannot be ignited, i.e., ignition degenerates and the particle burnout is limited to the kinetic-controlled regime without undergoing a thermal runway or transitioning to an external-diffusion-controlled regime. This finding is in contrast to the previous work by Soo~\textit{et al}.~\cite{Soo2018}, who found a minimum particle size for ignition, using a generic combustion model of metal particles that did not consider the effect of the oxide layer described by a parabolic rate law.\\

As shown in Fig.~\ref{Fig7}(a), $T_\mathrm{ign}$ resulting from the unsteady model, which considers the thickening of the oxide layer over time, is significantly greater than the steady-state estimation and plateaus at approximately $\SI{1080}{\kelvin}$ for $\delta_0 \lesssim 0.003$, as indicated by the gray vertical line in Fig~\ref{Fig7}(a). To understand this plateau behavior, the transient process of oxide-layer growth must be considered.\\

\begin{figure}[h!]
\centerline{\includegraphics[width=1.0\textwidth]{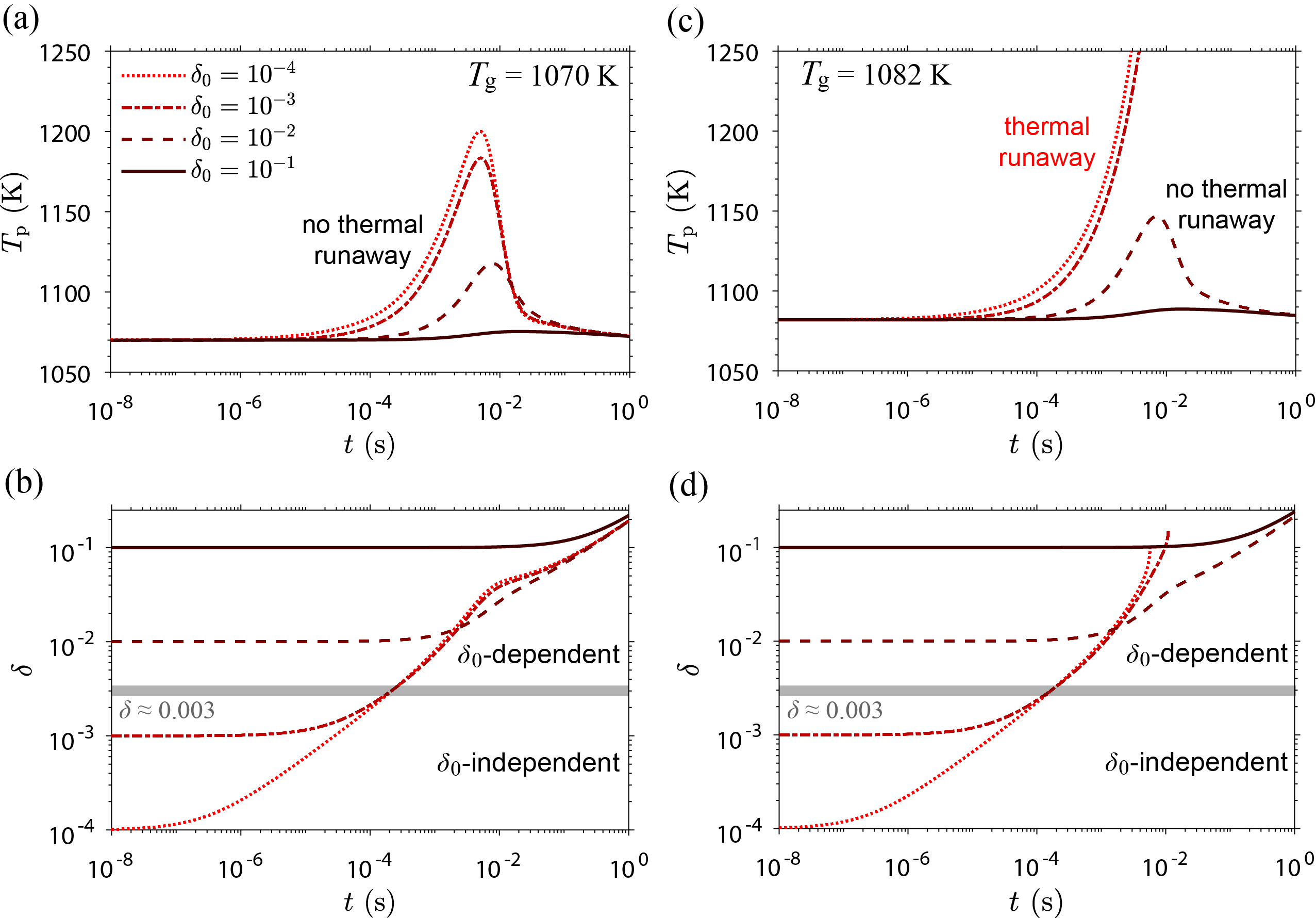}}
		\caption{The time histories of particle temperature, $T_\mathrm{p}$ [(a) and (c)] and the ratio between oxide layer thickness and particle radius, $\delta$ [(b) and (d)], for an isolated iron particle in air with $d_\mathrm{p,0}=\SI{20}{\micro\meter}$ and four different initial oxide thickness ratios, i.e., $\delta_0=10^{-4}$ (dotted), $\delta_0=10^{-3}$ (dash-dotted), $\delta_0=10^{-2}$ (dashed), and $\delta_0=10^{-1}$ (solid), subjected to two gas temperatures, i.e., $T_\mathrm{g} = \SI{1070}{\kelvin}$ for (a) and (b), and $T_\mathrm{g} = \SI{1082}{\kelvin}$ for (c) and (d).}
	\label{Fig8}
\end{figure}

The time histories of particle temperature and oxide-layer growth resulting from the unsteady model for an isolated particle in air with $d_\mathrm{p,0}=\SI{20}{\micro\meter}$ and different initial oxide-thickness ratios subjected to two gas temperatures ($T_\mathrm{g}=\SI{1070}{\kelvin}$ and $T_\mathrm{g}=\SI{1082}{\kelvin}$) are shown in Fig.~\ref{Fig8}. The subfigures (a) and (b) correspond to the cases with a gas temperature $T_\mathrm{g}=\SI{1070}{\kelvin}$ that is slightly below the plateau value of $T_\mathrm{ign}=\SI{1080}{\kelvin}$ marked in Fig.~\ref{Fig7}. None of these cases, with various $\delta_0$, results in a thermal runaway.  As shown in Fig.~\ref{Fig8}(b), $\delta$ increases the most rapidly in the case with $\delta_0 = 10^{-4}$ and collapses onto the curve of $\delta_0 = 10^{-3}$ upon reaching $\delta \approx 0.003$ (indicated by the gray horizontal line) around $t=\SI{1e-4}{\second}$. Figures~\ref{Fig8}(c) and (d) are for the cases with $T_\mathrm{g}=\SI{1082}{\kelvin}$, which is slightly greater than the plateau $T_\mathrm{ign}$. The cases with $\delta_0 = 10^{-4}$ and $\delta_0 = 10^{-3}$ result in a thermal runaway while the other two cases, with greater values of $\delta_0$, do not. Note that the curves of $\delta$ for the cases with $\delta_0 = 10^{-4}$ and $\delta_0 = 10^{-3}$ shown in Fig.~\ref{Fig8}(d) are cut off because the calculations were stopped once the melting point of \ch{FeO} is reached. Again, the $\delta$ growth curves for the cases with $\delta_0 = 10^{-4}$ and $\delta_0 = 10^{-3}$ collapse upon reaching $\delta \approx 0.003$ around $t=\SI{1e-4}{\second}$. By comparing the cases with  $T_\mathrm{g}=\SI{1070}{\kelvin}$ and  $T_\mathrm{g}=\SI{1082}{\kelvin}$, the characteristic time scale of thermal runaway is identified to be greater than $\SI{1e-3}{\second}$. The order of magnitude of this time scale suggests that, for increasingly thin initial oxide layers below $\delta_0 \approx 0.003$, a rapid oxide-layer growth up to a time scale of $\SI{1e-4}{\second}$ has no effect on the later processes determining whether a thermal runaway occurs. For sufficiently thin initial oxide layers, $T_\mathrm{ign}$ is thus independent of $\delta_0$.\\

\subsection{\label{Sect5_2}Effect of inert gas species}

\begin{figure}[h!]
\centerline{\includegraphics[width=0.55\textwidth]{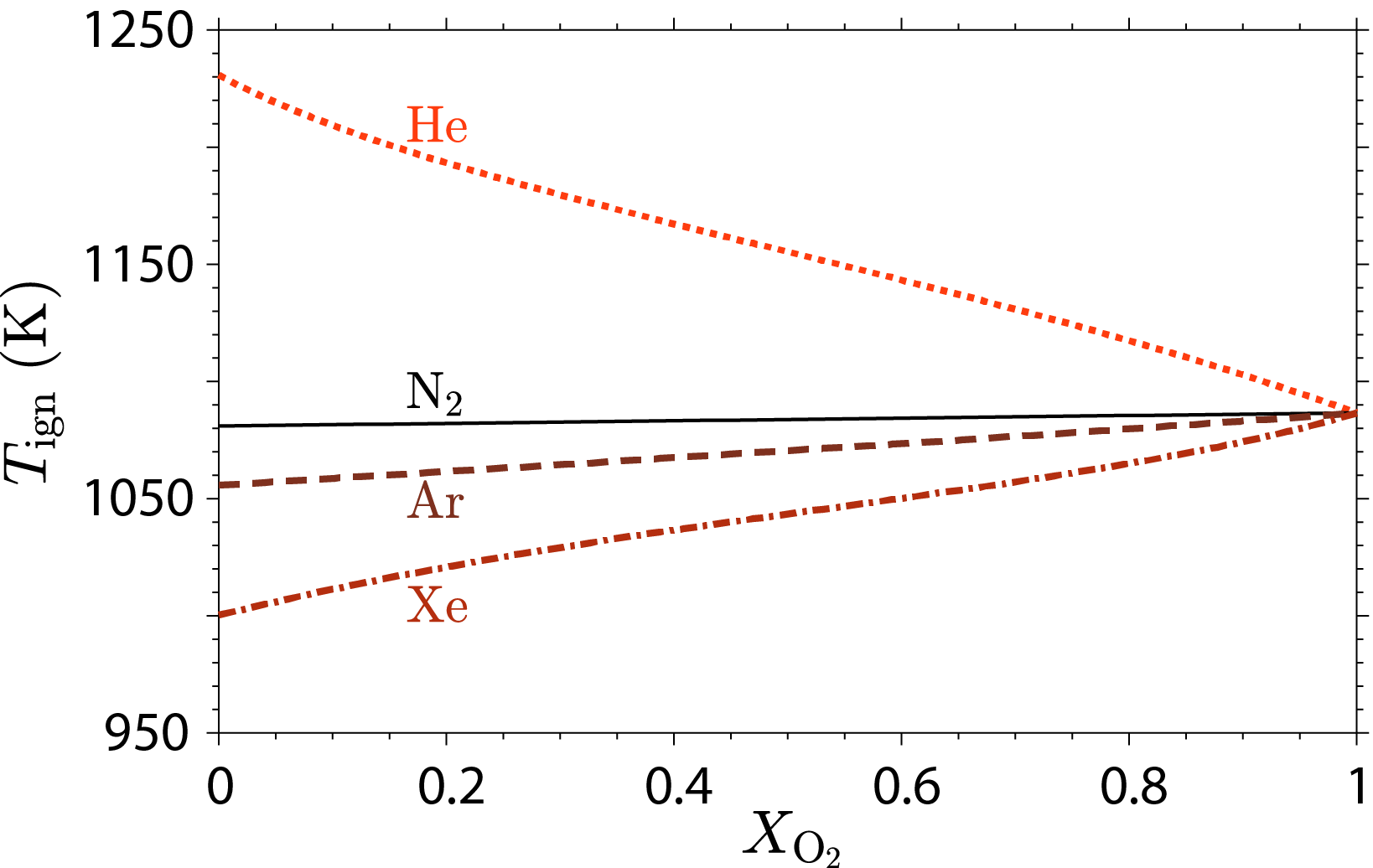}}
		\caption{The ignition temperature $T_\mathrm{ign}$ of an isolated iron particle with $\delta_0=10^{-3}$ as a function of mole fraction of oxygen $X_{\ch{O2}}$ in a bulk gas mixture with different inert species, i.e., \ch{He} (dotted), \ch{N2} (solid), \ch{Ar} (dashed), and \ch{Xe} (dash-dotted).}
	\label{Fig9}
\end{figure}

The current model considers the fact that the kinetics of iron oxidation is controlled by solid-phase diffusion and is independent of gas-phase \ch{O2} concentration (at least, for $p_{\ch{O2}} > \SI{4e-4}{atm}$). This means that, as the gaseous composition is varied, the ignition characteristics of an iron particle are only influenced by the change in the thermal conductivity of the ambient gas. The $T_\mathrm{ign}$ resulting from the unsteady model for an isolated particle with $\delta_0=10^{-3}$, in a gas mixture with \ch{O2} and different inert species, is plotted as a function of the mole fraction of \ch{O2} in Fig.~\ref{Fig9}. The result of $T_\mathrm{ign}$ for the case with \ch{N2} (plotted as the solid curve) does not change significantly as $X_{\ch{O2}}$ varies, owing to the fact that the thermal conductivities of diatomic gases \ch{O2} and \ch{N2} are very close. For the case with \ch{He}, the resulting $T_\mathrm{ign}$ increases with a decrease in $X_{\ch{O2}}$. This trend is due to the fact that \ch{He} has a higher thermal conductivity than that of \ch{O2}. A greater mole fraction of \ch{He} thus enhances the rate of conductive heat loss from the particle to the gas mixture, resulting in a greater $T_\mathrm{ign}$. On the other hand, the $T_\mathrm{ign}$ for the cases with \ch{Ar} and \ch{Xe} decrease significantly as oxygen mole fraction decreases, since \ch{Ar} and \ch{Xe} have an increasingly low thermal conductivity. A reduced thermal conductivity of the ambient gas has a heat preserving effect on a reacting particle, thereby, facilitating thermal runaway.\\

\subsection{\label{Sect5_3}Effect of radiative heat transfer}

\begin{figure}
\centerline{\includegraphics[width=1.0\textwidth]{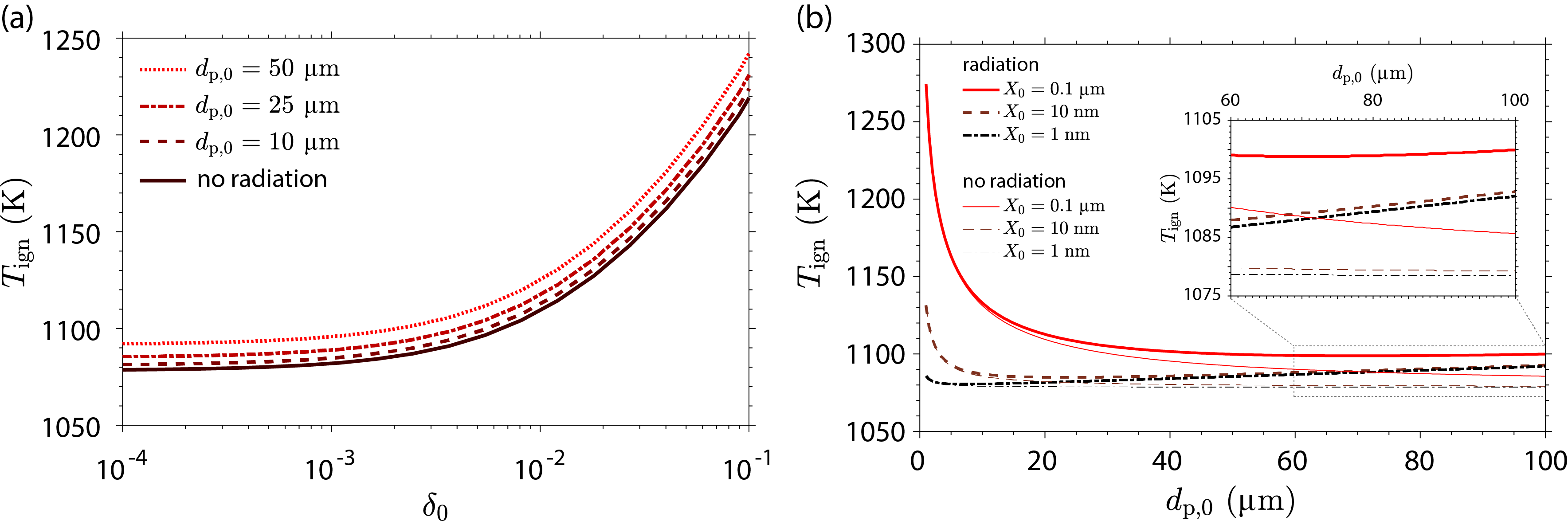}}
		\caption{The ignition temperature, $T_\mathrm{ign}$, of an isolated iron particle of (a) different initial sizes, i.e., $d_\mathrm{p,0}=\SI{50}{\micro\meter}$ (dotted), $d_\mathrm{p,0}=\SI{25}{\micro\meter}$ (dash-dotted), and $d_\mathrm{p,0}=\SI{10}{\micro\meter}$ (dashed), with radiative heat transfer to the ambient air and without radiative heat transfer (solid) plotted as a function of $\delta_0$, and (b) different initial oxide layer thicknesses, i.e., $X_0=\SI{0.1}{\micro\meter}$ (solid), $X_0=\SI{10}{\nano\meter}$ (dashed), and $X_0=\SI{1}{\nano\meter}$ (dash-dotted), with (thick curves) and without (thin curves) radiative heat transfer plotted as a function of $d_\mathrm{p,0}$. The inset in (b) is a zoom-in view showing the resulting trends of $T_\mathrm{ign}$ for relatively large particles.}
	\label{Fig10}
\end{figure}

If the heat loss from a particle to the surrounding gas via radiative heat transfer is considered, the criterion of thermal runaway (formulated as Eqs.~\ref{Eq19} and \ref{Eq20}) is augmented as follows:
\begin{equation}
\begin{split}
     & q_{\ch{FeO}} \frac{\mathrm{d} m_{\ch{FeO}}}{\mathrm{d} t} + q_{\ch{Fe3O4}} \frac{\mathrm{d} m_{\ch{Fe3O4}}}{\mathrm{d} t}  >  A_\mathrm{p} h_\mathrm{p} (T_\mathrm{p}-T_\mathrm{g}) + A_\mathrm{p} \sigma \epsilon ({T_\mathrm{p}}^4-{T_\mathrm{g}}^4)\\[3pt]
     & \tilde{q}_{\ch{FeO}}  \; \mathrm{Exp} \left( \frac{-T_\mathrm{a,FeO}}{T_\mathrm{p}} \right) \frac{{(1-\delta \delta_{\ch{Fe3O4}})}^2 {\color{red}{r_\mathrm{p}}}}{\delta -\delta \delta_{\ch{Fe3O4}}} + \tilde{q}_{\ch{Fe3O4}}  \; \mathrm{Exp} \left( \frac{-T_\mathrm{a,\ch{Fe3O4}}}{T_\mathrm{p}} \right) \frac{{\color{red}{r_\mathrm{p}}}}{\delta \delta_{\ch{Fe3O4}}} > \lambda_\mathrm{g} (T_\mathrm{p} - T_\mathrm{g}) {\color{red}{r_\mathrm{p}}} + \sigma \epsilon ({T_\mathrm{p}}^4-{T_\mathrm{g}}^4) {\color{red} {r_\mathrm{p}}^{2}}.
\end{split}
\label{Eq21}
\end{equation}
Note that the radiative heat loss rate scales quadratically with particle size while the energy release rates and conductive heat loss rate scale linearly with $r_\mathrm{p}$. The heat removal effect due to radiation is thus enhanced for increasingly large particles. This effect is reflected by the results of $T_\mathrm{ign}$ as a function of $\delta_0$, for the cases with various initial particle sizes, and $T_\mathrm{ign}$ as a function of $d_\mathrm{p,0}$, for the cases with different initial oxide layer thicknesses, as shown in Figs.~\ref{Fig10}(a) and (b), respectively. In Fig.~\ref{Fig10}(a), the resulting $T_\mathrm{ign}$ for the cases with radiative heat loss are greater than the  $T_\mathrm{ign}$ for the case neglecting radiation (the solid curve), and increase with an increasingly large $d_\mathrm{p,0}$ for all values of $\delta_0$. For a fixed initial oxide layer thickness $X_0$ and relatively small particle sizes, as shown in Fig.~\ref{Fig10}(b), $T_\mathrm{ign}$ decreases as $d_\mathrm{p,0}$ increases. The results for the cases with (thick curves in Fig.~\ref{Fig10}(b)) and without (thin curves) radiative heat loss are indistinguishable below approximately $d_\mathrm{p,0}=\SI{10}{\micro\meter}$. For relatively large particles, e.g., $d_\mathrm{p,0} \geq \SI{60}{\micro\meter}$ as shown in the inset of Fig.~\ref{Fig10}(b), $T_\mathrm{ign}$ resulting from the cases with radiative heat loss slightly increases with an increase in $d_\mathrm{p,0}$ while the $T_\mathrm{ign}$ without radiation effect slightly decrease or plateau. As demonstrated by the results in Fig.~\ref{Fig10}, the difference in $T_\mathrm{ign}$ between the cases with and without radiative heat loss is less than \SI{20}{\kelvin}, i.e., a relative discrepancy less than $2\%$, suggesting that radiative heat loss has a minor effect on the ignition characteristics of fine iron particles.

\subsection{\label{Sect5_4}Collective heating effect in a suspension of particles}

\begin{figure}[h!]
\centerline{\includegraphics[width=1.0\textwidth]{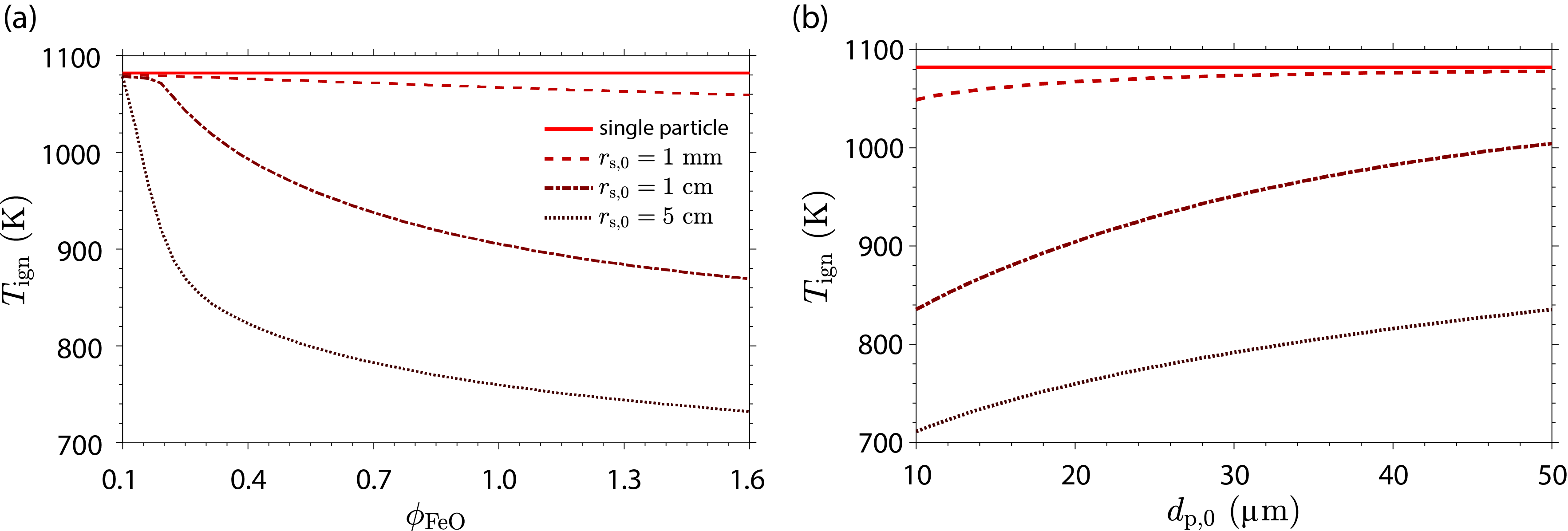}}
		\caption{The ignition temperature, $T_\mathrm{ign}$, of a suspension of iron particles in air with different initial suspension radii, i.e., $r_\mathrm{s,0}=\SI{1}{\milli\meter}$ (dashed), $r_\mathrm{s,0}=\SI{1}{\centi\meter}$ (dash-dotted), and $r_\mathrm{s,0}=\SI{5}{\centi\meter}$ (dotted), and $\delta_0=10^{-3}$ (a) as a function of fuel equivalence ratio (assuming \ch{FeO} as the only product) with $d_\mathrm{p,0}=\SI{20}{\micro\meter}$ and (b) as a function of initial particle diameter with $\phi_{\ch{FeO}}=1$. The value of $T_\mathrm{ign}$ for an isolated particle (with $\delta_0=10^{-3}$) in air is marked by the horizontal solid line.}
	\label{Fig11}
\end{figure}

The results of $T_\mathrm{ign}$ for a spherical suspension of iron particles (with $\delta_0=10^{-3}$ and a fixed initial particle size of $d_\mathrm{p,0}=\SI{20}{\micro\meter}$) in air are plotted as a function of $\phi_{\ch{FeO}}$, i.e., the fuel equivalence ratio considering \ch{FeO} as the only oxide product in Fig.~\ref{Fig11}(a). The calculations were performed for a constant gas composition $X_{\ch{O2}}$:$X_{\ch{N2}}$=$0.21$:$0.79$ while $\phi_{\ch{FeO}}$ is varied by changing the number of particles inside the suspension according to Eqs.~\ref{Eq11} and \ref{Eq12}. As the initial radius of a spherical suspension increases from $r_\mathrm{s,0}=\SI{1}{\milli\meter}$ (dashed curve) to \SI{5}{\centi\meter} (dotted curve), the resulting $T_\mathrm{ign}$ decreases from approximately \SI{1075}{\kelvin}, which is slightly below the $T_\mathrm{ign}$ for an isolated particle (indicated by the solid horizontal line), to values below \SI{800}{\kelvin}. This reduction in $T_\mathrm{ign}$ is due to the collective effect of energy released by multiple particles suspended within a finite-sized volume that is losing heat to the surrounding via its external surface~\cite{Soo2018}. Analogous to an isolated particle, the conductive heat loss rate of a spherical suspension scales linearly with suspension radius as $A_\mathrm{s}~\propto~{r_\mathrm{s}}^2$ and $h_\mathrm{s}~\propto~1/r_\mathrm{s}$. The rate of energy release from the reacting particles in a suspension scales with $N_\mathrm{p}$ that is proportional to ${r_\mathrm{s}}^3$. The increase in energy release rate thus surpasses the increase in heat loss rate for an increasingly large suspension. This analysis suggests that, as a result of the collective effect, the critical ignition temperature of iron particles within a flame may be lower than that for an isolated particle. Future experimental effort is required to quantitatively probe the collective heating effect on flame propagation in suspensions of iron particles.\\

Another remark on the result shown in Fig.~\ref{Fig11}(a) is that $T_\mathrm{ign}$ monotonically decreases as $\phi_{\ch{FeO}}$ increases. As the kinetic oxidation rate is independent of the \ch{O2} concentration of the suspension, the overall energy release rate is not affected by a slight depletion of \ch{O2} prior to transitioning to an external-diffusion-controlled combustion. Thus, the ignition propensity of a suspension is enhanced by an increase in volumetric concentration of iron particles, regardless of the corresponding equivalence ratio. As $\phi_{\ch{FeO}}$ decreases to very small values, the volumetric number density of iron particles approaches zero and, thus, the resulting $T_\mathrm{ign}$ of a suspension reverts back to the value of an isolated particle.\\

Although $T_\mathrm{ign}$ of an isolated iron particle is independent of initial particle size for a fixed value of $\delta_0$ (as discussed in Sect.~\ref{Sect5_1}), Fig.~\ref{Fig11}(b) shows that $T_\mathrm{ign}$ of a suspension increases with $d_\mathrm{p,0}$ of individual particles for a fixed equivalence ratio $\phi_{\ch{FeO}}=1$. For a fixed iron concentration in a suspension, the smaller the particles are, the larger internal surface area is available to reaction, while the external surface area of the suspension, for heat loss into the surrounding, remains the same. Thus, smaller particles give rise to a more pronounced collective effect on enhancing the ignition propensity of a suspension. A similar enhancement due to the collective effect for smaller particles has also been identified in the ignition analysis for agglomerates of metal particles~\cite{Mi2013,Soo2018}.

\subsection{\label{Sect5_5}Transition from kinetic- to diffusion-controlled combustion}

\begin{figure}[h!]
\centerline{\includegraphics[width=0.55\textwidth]{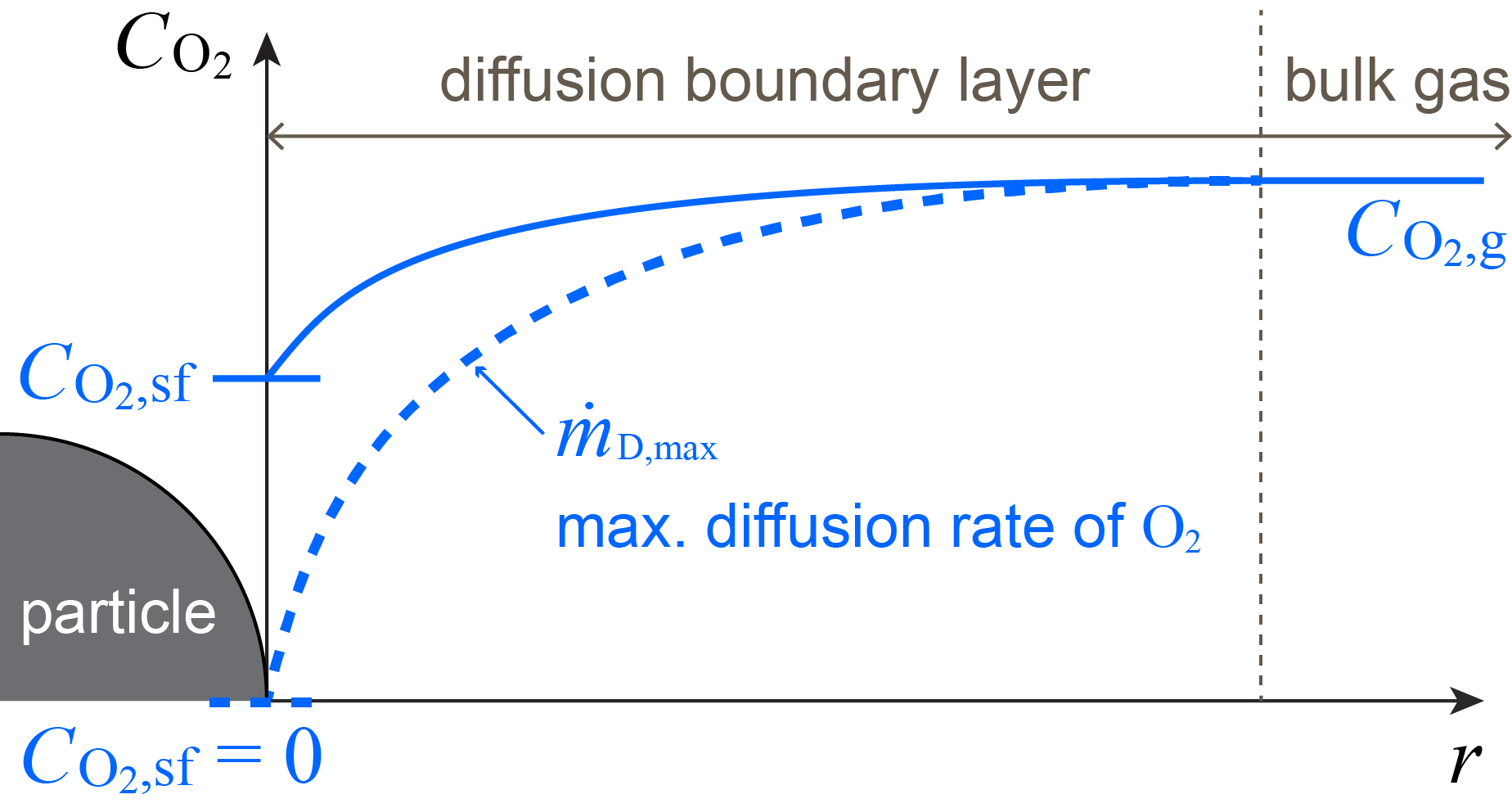}}
		\caption{Schematic illustrating the diffusion of \ch{O2} from the bulk gas to the particle surface. The dashed curve corresponds to the case of the maximum diffusion rate at a given \ch{O2} concentration in the bulk gas.}
	\label{Fig12}
\end{figure}

Although the current model considering a completely kinetic-controlled reaction rate is sufficient to predict the ignition temperature, for temperatures greater than $T_\mathrm{ign}$, it can be further developed to estimate the conditions under which an iron particle transitions from a kinetic-controlled combustion to an external-diffusion-controlled combustion. This further development is based on the generic $k$-$\beta$ paradigm~\cite{Soo2018} of particle reaction that considers the interplay between the condensed-phase (or heterogeneous) oxidation kinetics and the transport of oxidizer from the ambient gas to the particle surface. The kinetic consumption rate of \ch{O2}, denoted as $\dot{m}_\mathrm{R}$, is governed by the parabolic rate laws of iron oxidation as described by Eq.~\ref{Eq8}. This kinetic rate is approximated to be independent of the gaseous \ch{O2} concentration at the particle surface, $C_{\ch{O2}\mathrm{,sf}}$, for any $C_{\ch{O2}\mathrm{,sf}}>0$, as justified in Sect.~\ref{Sect2_1}. The rate of \ch{O2} mass transport, $\dot{m}_\mathrm{D}$, from the ambient gas to the particle surface (as illustrated in Fig.~\ref{Fig12}) can be estimated as follows,
\begin{equation}
    \dot{m}_\mathrm{D} = A_\mathrm{p} \beta_\mathrm{p} (C_{\ch{O2}\mathrm{,g}}-C_{\ch{O2}\mathrm{,sf}})
\end{equation}
where $C_{\ch{O2}\mathrm{,g}}$ and $C_{\ch{O2}\mathrm{,sf}}$ are the concentrations of \ch{O2} in the bulk gas and at the particle surface, respectively. The diffusive mass transfer coefficient $\beta_\mathrm{p}$ is expressed as:
\begin{equation}
    \beta_\mathrm{p} = \frac{\mathrm{Sh} D_{\ch{O2}}}{d_\mathrm{p}}.
\end{equation}
The Sherwood number ($\mathrm{Sh}$) for a spherical particle can also be calculated using the Fr\"{o}ssling correlation:
\begin{equation}
\mathrm{Sh}=2+0.552 \mathrm{Re}^{\frac{1}{2}} \mathrm{Sc}^{\frac{1}{3}}.
\end{equation}
Under the assumption of a quiescent gas medium, Sherwood number equals to two, thus, $\beta_\mathrm{p}=D_{\ch{O2}}/r_\mathrm{p}$. The diffusivity, $D_{\ch{O2}}$, through this diffusion boundary layer around a particle is estimated based on the mixture-averaged thermal conductivity calculated via Eq.~\ref{Eq10} assuming a unity Lewis number. The maximum \ch{O2} diffusion rate corresponds to the case wherein $C_{\ch{O2}\mathrm{,sf}}=0$ as illustrated by the dashed curve in Fig.~\ref{Fig12} and can be calculated as:
\begin{equation}
  \dot{m}_\mathrm{D,max} = 4 \pi r_\mathrm{p} D_{\ch{O2}} C_{\ch{O2}\mathrm{,g}},
  \label{Eq24}
\end{equation}
where $D_{\ch{O2}}$ is a function of gas composition and particle surface temperature, $T_\mathrm{sf}$, estimated via the two-third law based on $T_\mathrm{p}$ and $T_\mathrm{g}$.\\

\begin{figure}[h!]
\centerline{\includegraphics[width=1.0\textwidth]{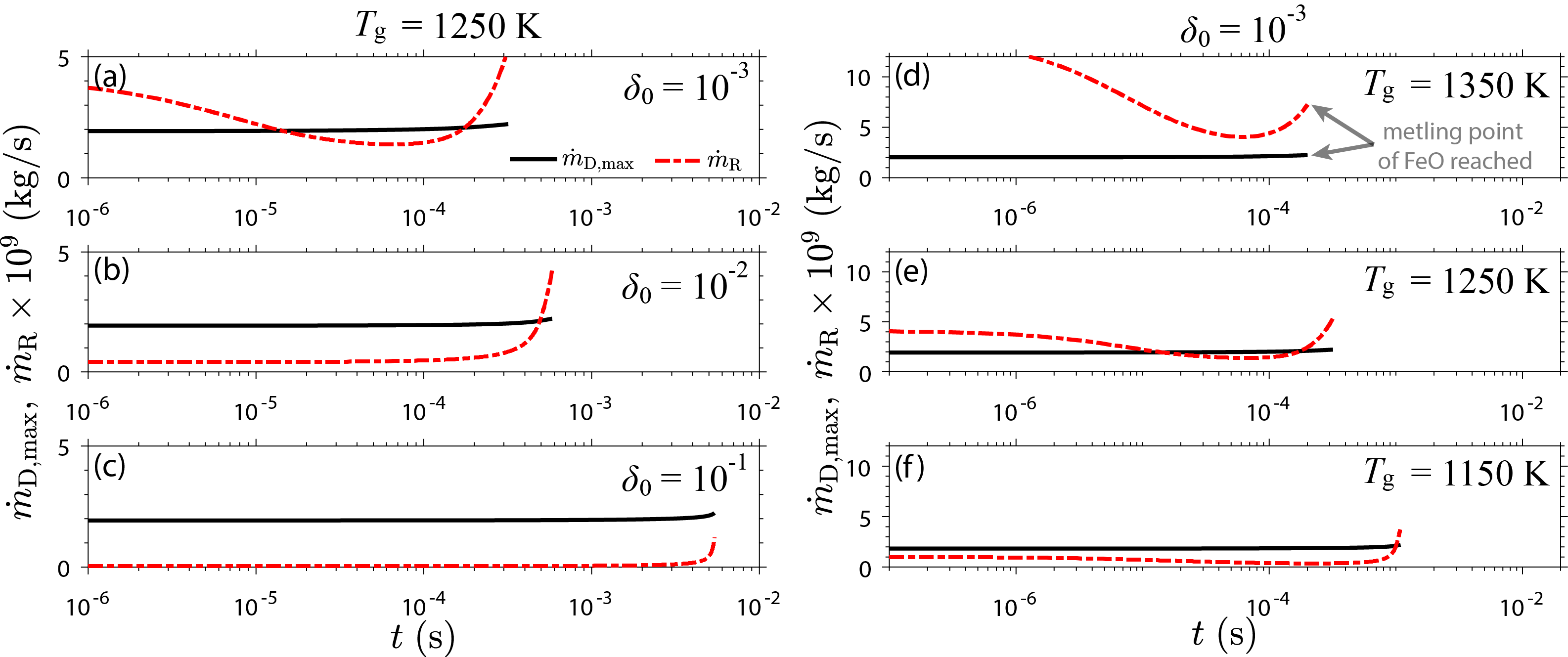}}
		\caption{Comparison between the kinetic consumption rate, $\dot{m}_\mathrm{R}$, and the maximum external diffusion rate, $\dot{m}_\mathrm{D,max}$, of \ch{O2} in air for the cases of an initially \SI{20}{\micro\meter}-sized particle with (a-c) a fixed gas temperature of $T_\mathrm{g}=\SI{1250}{\kelvin}$ and increasingly large initial oxide thickness ratios and (d-f) a fixed $\delta_0=10^{-3}$ and increasingly high gas temperatures.}
	\label{Fig13}
\end{figure}

For $C_{\ch{O2}\mathrm{,sf}}>0$, the kinetic rate is less than the maximum rate of \ch{O2} diffusion, i.e., $\dot{m}_\mathrm{R} < \dot{m}_\mathrm{D,max}$, and the \ch{O2} consumption rate of the particle, $\dot{m}_{\ch{O2}}$, is completely controlled by the solid-phase kinetics. When $C_{\ch{O2}\mathrm{,sf}}=0$, the maximum \ch{O2} diffusion rate is reached such that $\dot{m}_{\ch{O2}}$ must be equal to $\dot{m}_\mathrm{D,max}$. Considering the nature of solid-phase iron oxidation, the generic $k$-$\beta$ model can thus be specialized to a \textit{switch}-type model between the kinetic- and external-diffusion-controlled oxidation rates as:
\begin{equation}
    \begin{cases}
      \dot{m}_{\ch{O2}} = \dot{m}_\mathrm{R} & \mathrm{if}\; \dot{m}_\mathrm{R} < \dot{m}_\mathrm{D,max}\\
      \dot{m}_{\ch{O2}} = \dot{m}_\mathrm{D,max} & \mathrm{otherwise}
    \end{cases}\;,
    \label{EqNewKBeta}
\end{equation}
to more accurately describe the transitions between the particle combustion regimes at temperatures greater than $T_\mathrm{ign}$. Once a particle transitions into external-diffusion-limited combustion, the total reaction rate of \ch{O2}, $\dot{m}_{\ch{O2}}$, is partitioned into the formation rates of \ch{FeO} and \ch{Fe3O4} based on the ratio between the kinetic rates (as shown in Fig.~\ref{Fig2}) of \ch{FeO} and \ch{Fe3O4} formation at the corresponding $T_\mathrm{p}$, $X_{\ch{FeO}}$, and $X_{\ch{Fe3O4}}$. Figure~\ref{Fig13} shows the comparison between $\dot{m}_\mathrm{R}$ (dash-dotted curve) and $\dot{m}_\mathrm{D,max}$ (solid curve) resulting from the unsteady model for an isolated particle of $d_\mathrm{p,0}=\SI{20}{\micro\meter}$ in air. When $\dot{m}_\mathrm{R}$ exceeds $\dot{m}_\mathrm{D,max}$, it indicates the transition of a particle from a kinetic-controlled to an external-diffusion-controlled combustion. The results of $\dot{m}_\mathrm{R}$ and $\dot{m}_\mathrm{D,max}$ stop when $T_\mathrm{p}$ reaches the melting point (m.p.) of \ch{FeO}.\\

\begin{figure}[h!]
\centerline{\includegraphics[width=0.6\textwidth]{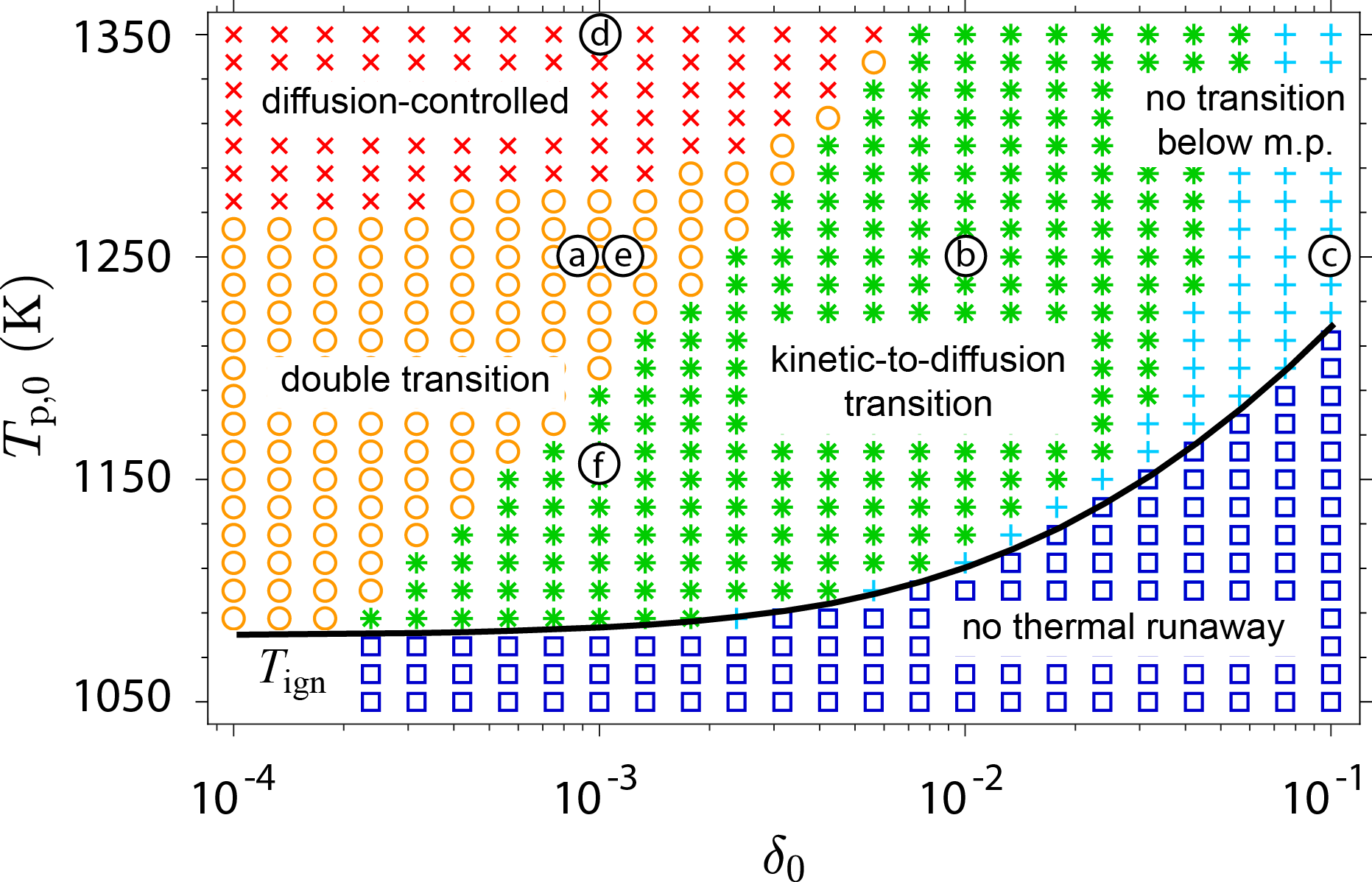}}
		\caption{A map in the parametric space of initial particle temperature $T_\mathrm{p,0}$ ($=T_\mathrm{g}$) and $\delta_0$ summarizing different behaviors of transition (prior to the melting point of \ch{FeO}) between particle combustion regimes that are identified via the comparison between $\dot{m}_\mathrm{R}$ and $\dot{m}_\mathrm{D,max}$ for the case with $d_\mathrm{p,0}=\SI{20}{\micro\meter}$. The different transition behaviors are marked as follows: {\color{red}$\times$} Completely external \ch{O2}-diffusion-controlled ($\dot{m}_\mathrm{R}>\dot{m}_\mathrm{D,max}$, e.g., Fig.~\ref{Fig13}(d)); {\color{orange}$\circ$} transition from an initially diffusion- to kinetic-controlled regime followed by a transition from kinetic-  to diffusion-controlled regime ($\dot{m}_\mathrm{R}>\dot{m}_\mathrm{D,max} \to \dot{m}_\mathrm{R}<\dot{m}_\mathrm{D,max} \to \dot{m}_\mathrm{R}>\dot{m}_\mathrm{D,max}$, e.g., Figs.~\ref{Fig13}(a) and (e)); {\color{green}$\ast$} transition from kinetic- to diffusion-controlled regime ($\dot{m}_\mathrm{R}<\dot{m}_\mathrm{D,max} \to \dot{m}_\mathrm{R}>\dot{m}_\mathrm{D,max}$, e.g., Figs.~\ref{Fig13}(b) and (f)); {\color{cyan}$+$} completely kinetic-controlled, no transition below the melting point of \ch{FeO} ($\dot{m}_\mathrm{R}<\dot{m}_\mathrm{D,max}$, e.g., Fig.~\ref{Fig13}(c)).}
	\label{Fig14}
\end{figure}

The results for the cases with a fixed gas temperature of $T_\mathrm{g}=\SI{1250}{\kelvin}$ and various initial oxide thicknesses are plotted in Fig.~\ref{Fig13}(a)-(c), and for the cases with a fixed initial oxide thickness ratio of $\delta_0=10^{-3}$ and various gas temperatures are plotted in Fig.~\ref{Fig13}(d)-(f). For a relatively thin oxide layer of $\delta_0=10^{-3}$ and a moderately high gas temperature of \SI{1250}{\kelvin}, as shown in Fig.~\ref{Fig13}(a) and (e), $\dot{m}_\mathrm{R}$ is initially greater than $\dot{m}_\mathrm{D,max}$ for a very short period of time and then decreases below $\dot{m}_\mathrm{D,max}$ around $t=\SI{10}{\micro\second}$ due to the growth of the oxide layer slowing down the reaction rate. Over this short period of $\dot{m}_\mathrm{R}>\dot{m}_\mathrm{D,max}$, the particle reaction rate is external-diffusion controlled. After a short oxidation phase, the oxide layer grows sufficiently that the reaction rate slows and a transition to kinetic control is observed. At a later time around $t=\SI{2e-4}{\second}$, $\dot{m}_\mathrm{R}$ again becomes greater than $\dot{m}_\mathrm{D,max}$. This crossover marks the transition of the particle from a kinetic-controlled combustion to an external-diffusion-controlled combustion prior to reaching the melting point of \ch{FeO}. This \textit{double transition} behavior is identified in the region of thin initial oxide layer and moderately high initial particle temperature (which equals to gas temperature) as marked by the orange circles {\color{orange}$\circ$} in Fig.~\ref{Fig14}.\\

For a fixed gas temperature at \SI{1250}{\kelvin} with a larger initial oxide thickness ratio $\delta_0=10^{-2}$, the particle initially burns under a kinetic-controlled regime ($\dot{m}_\mathrm{R}<\dot{m}_\mathrm{D,max}$) and transitions to a diffusion-controlled combustion shortly prior to reaching the melting point of \ch{FeO} as shown in Fig.~\ref{Fig13}(b). A similar transition behavior is found for the case with a smaller initial oxide ratio $\delta_0=10^{-3}$ and a lower gas temperature at \SI{1150}{\kelvin} as shown in Fig.~\ref{Fig13}(f). In Fig.~\ref{Fig14}, the green asterisks {\color{green}$\ast$} mark the cases with a kinetic-to-diffusion transition below the melting point of \ch{FeO}.\\

For cases with a sufficiently thin initial oxide layer and high gas temperature, e.g., $\delta_0=10^{-3}$ and $T_\mathrm{g}=\SI{1350}{\kelvin}$ as shown in Fig.~\ref{Fig13}(d), $\dot{m}_\mathrm{R}$ remains greater than $\dot{m}_\mathrm{D,max}$ throughout the combustion process prior to the \ch{FeO} melting point, suggesting that the thermal runaway process is likely to be completely controlled by the external diffusion of \ch{O2}. This scenario is identified for the cases with a thin initial oxide thickness and high gas temperature as marked by the red crosses {\color{red}$\times$} in Fig.~\ref{Fig14}. This finding implies that, if a particle is abruptly heated to a sufficiently high temperature without a significant growth of oxide layer (e.g., by laser heating in Ref.~\cite{Ning2021}), its ignition characteristics may not be governed by the kinetics of solid-phase iron oxidation at all.\\

In the case with $T_\mathrm{g}=\SI{1250}{\kelvin}$ and an even thicker initial oxide layer of $\delta_0=10^{-1}$, $\dot{m}_\mathrm{R}$ remains smaller than $\dot{m}_\mathrm{D,max}$ throughout the entire thermal runaway process prior to reaching the \ch{FeO} melting point as shown in Fig.~\ref{Fig13}(c). Thus, no transition from kinetic- to diffusion-controlled combustion occurs before the particle starts to melt. The region in the $T_\mathrm{p,0}$-$\delta_0$ parametric space marked by the cyan plus symbols {\color{cyan}$+$} in Fig.~\ref{Fig14} corresponds to the cases without a transition below the melting point.\\

\subsection{\label{Sect5_7}The characteristics of ignition delay time, $\tau_\mathrm{ign}$}

For isolated particles with heat loss to the surrounding, ignition delay times, $\tau_\mathrm{ign}$, can be determined for the cases with a gas temperature (and initial particle temperature) equal to or greater than the ignition temperature, i.e., $T_\mathrm{g}=T_\mathrm{p,0} \geq T_\mathrm{ign}$, using the augmented reaction-rate model, Eq.~\ref{EqNewKBeta}. For metal particles, since the increase of temperature is interrupted by phase changes, ignition delay time cannot be conveniently measured based on the moment of maximum temperature increase as commonly defined for homogeneous reactive mixtures. In this study, $\tau_\mathrm{ign}$ is measured as the time elapsed from the beginning of each calculation with $T_\mathrm{g}=T_\mathrm{p,0}$ to the moment when $T_\mathrm{p}$ reaches the melting point of \ch{FeO}, i.e., \SI{1640}{\kelvin}. This quantity can be measured and compared consistently in future studies. The effects of initial oxide-layer thickness and particle size on $\tau_\mathrm{ign}$ of isolated particles are first examined in Sects.~\ref{Sect5_7_1} and~\ref{Sect5_7_2}, respectively. In order to better probe the dependence of $\tau_\mathrm{ign}$ on the reaction nature of iron particles, $\tau_\mathrm{ign}$ is also analyzed for adiabatic suspensions of particles in Sect.~\ref{Sect5_7_3}.

\subsubsection{\label{Sect5_7_1}Effect of initial oxide-layer thickness on $\tau_\mathrm{ign}$ of isolated particles}
\begin{figure}[h!]
\centerline{\includegraphics[width=1.0\textwidth]{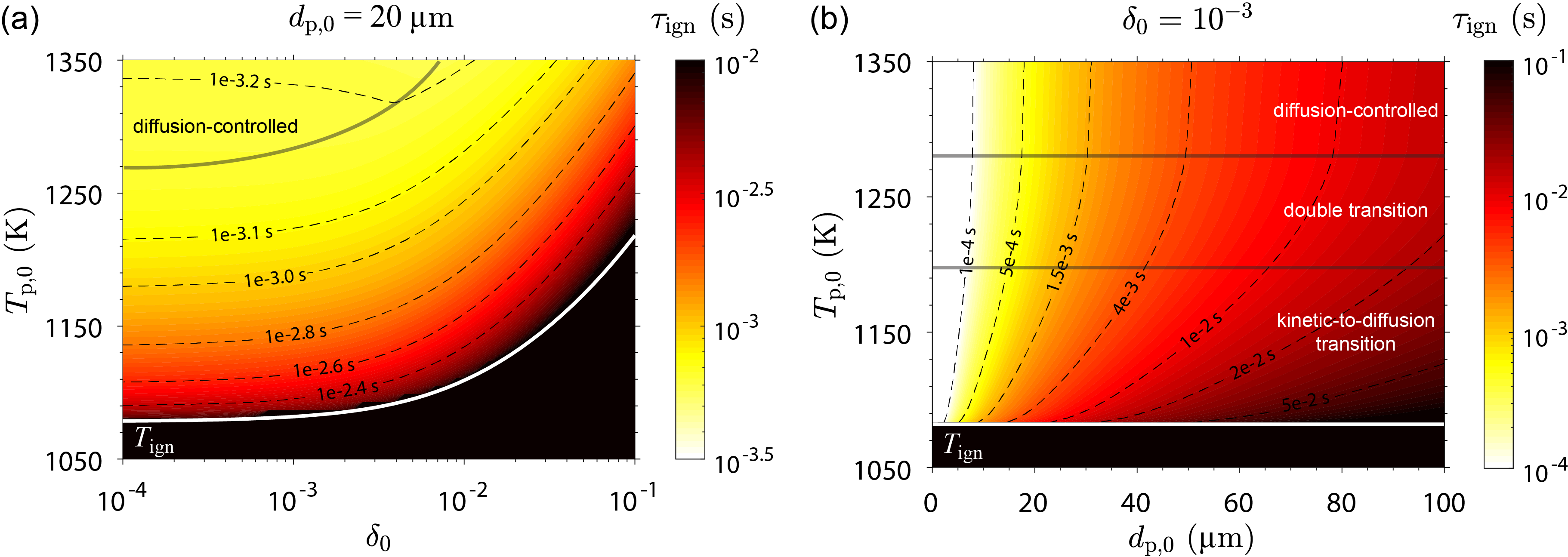}}
		\caption{Contour plots of ignition delay time, $\tau_\mathrm{ign}$, in the parametric spaces of (a) initial particle temperature $T_\mathrm{p,0}$ ($=T_\mathrm{g}$) and $\delta_0$ for a constant initial particle size of $d_\mathrm{p,0}=\SI{20}{\micro\metre}$ and (b) initial particle temperature and initial particle diameter for $\delta_0=10^{-3}$. The white curve in each plot indicates the ignition temperature, $T_\mathrm{ign}$. The dashed curves are the iso-contours of $\tau_\mathrm{ign}$. The gray curve in (a) indicates the boundary of the completely external-diffusion-controlled regime marked by {\color{red}$\times$} in Fig.~\ref{Fig14}. The gray horizontal lines in (b) indicate the boundaries (in $T_\mathrm{p,0}$) between the diffusion-controlled, double-transition, and kinetic-to-diffusion-controlled regimes, as identified in Fig.~\ref{Fig14}.}
	\label{FigTauIgn}
\end{figure}

For a selected initial particle size of $d_\mathrm{p,0}=\SI{20}{\micro\metre}$, the model prediction of $\tau_\mathrm{ign}$ for isolated iron particles are plotted as color contours in the parametric space of $T_\mathrm{p,0}$ and $\delta_0$, as shown in Fig.~\ref{FigTauIgn}(a). The predicted $T_\mathrm{ign}$ as a function of $\delta_0$ is plotted as the white curve on the color contours. The dashed curves indicate the iso-contours of $\tau_\mathrm{ign}$. The gray curve indicates the boundary of the completely external-diffusion-controlled region marked by {\color{red}$\times$} in Fig.~\ref{Fig14}. Figure~\ref{FigTauIgn}(a) shows that, outside the external-diffusion-controlled region, the iso-contours of $\tau_\mathrm{ign}$ follow qualitatively the same trend of $T_\mathrm{ign}$ as a function of $\delta_0$. For a fixed initial particle size, a particle with a greater $\delta_0$ requires a higher initial temperature to result in the same ignition delay time, reflecting the temperature dependence of kinetic-controlled combustion. Inside the diffusion-controlled region, $\tau_\mathrm{ign}$ barely varies with $T_\mathrm{p,0}$ or $\delta_0$.\\

In Fig.~\ref{FigTauIgn}(a), it is also of interest to note that the iso-contour of $\tau_\mathrm{ign}=10^{-3.2}~\mathrm{s}$ follows an opposite trend to those outside the diffusion region: For a fixed initial size, a particle with a greater $\delta_0$ requires a lower initial temperature to result in the same ignition delay time. The turning point of the iso-contour of $\tau_\mathrm{ign}=10^{-3.2}~\mathrm{s}$ is located on the boundary of the completely diffusion-controlled region, which is marked by the gray curve. This opposite trend is observed since, for a fixed particle size and temperature with an increasingly large $\delta_0$, the heat capacity (in $\SI{}{\joule\per\kelvin}$) of the particle slightly decreases, resulting in a stronger feedback to thermal runaway.\\

In Fig.~\ref{FigTauIgnCurves}(a), the predicted $\tau_\mathrm{ign}$ for the cases with $d_\mathrm{p,0}=\SI{20}{\micro\metre}$ and various values of $\delta_0$ are plotted as functions of the reciprocal of $T_\mathrm{p,0}$. The curves for the cases with $\delta_0=10^{-4}$ and $10^{-3}$ collapse onto a nearly horizontal line at high temperatures, corresponding to the diffusion-controlled regime identified in the $T_\mathrm{p,0}$-$\delta_0$ parametric space. As $T_\mathrm{p,0}$ approaches $T_\mathrm{ign}$, marked by the circles in Fig.~\ref{FigTauIgnCurves}(a), $\tau_\mathrm{ign}$ increases more rapidly than exponential due to an increasingly pronounced effect of heat loss from the particle to the surrounding.\\

\subsubsection{\label{Sect5_7_2}Effect of particle size on $\tau_\mathrm{ign}$ of isolated particles}

Figure~\ref{FigTauIgn}(b) shows the contour plot of $\tau_\mathrm{ign}$ in the parametric space of $T_\mathrm{p,0}$ and initial particle diameter, $d_\mathrm{p,0}$, for a fixed value of $\delta_0=10^{-3}$. Although $T_\mathrm{ign}$ is solely dependent on $\delta_0$, as indicated by the horizontal white line in Fig.~\ref{FigTauIgn}(b), $\tau_\mathrm{ign}$ significantly increases with particle size. For the cases with $T_\mathrm{p,0}$ slightly greater than $T_\mathrm{ign}$, $\tau_\mathrm{ign}$ increases over three orders of magnitude, i.e., from $10^{-4} \; \mathrm{s}$ to \SI{0.1}{\second}, as $d_\mathrm{p,0}$ increases from \SI{1}{\micro\metre} to \SI{100}{\micro\metre}. The horizontal gray lines mark the boundaries of different regimes of particle combustion in terms of $T_\mathrm{p,0}$, and show that the transition behavior between kinetic- and diffusion-controlled combustion regimes is independent of particle size. This independence is due to the fact that the kinetic rate of oxidation, $\dot{m}_\mathrm{R}$, and the maximum diffusion rate of \ch{O2} from the bulk gas, $\dot{m}_\mathrm{D,max}$, both linearly scale with particle size, as shown below:
\begin{equation}
\label{EqmRmD}
    \begin{split}
    & \dot{m}_\mathrm{R} = \tilde{\nu}_{\ch{O2}/\ch{FeO}}   \; \mathrm{Exp} \left( \frac{-T_\mathrm{a,FeO}}{T_\mathrm{p}} \right) \frac{{(1-\delta \delta_{\ch{Fe3O4}})}^2 {\color{red}{r_\mathrm{p}}} }{\delta -\delta \delta_{\ch{Fe3O4}}} + \tilde{\nu}_{\ch{O2}/\ch{Fe3O4}}  \; \mathrm{Exp} \left( \frac{-T_\mathrm{a,\ch{Fe3O4}}}{T_\mathrm{p}} \right) \frac{{\color{red}{r_\mathrm{p}}}}{\delta \delta_{\ch{Fe3O4}}} \\[5pt]
    & \dot{m}_\mathrm{D,max} = 4 \pi {\color{red}{r_\mathrm{p}}} D_{\ch{O2}} C_{\ch{O2}\mathrm{,g}},
  \end{split}  
\end{equation}
where $\tilde{\nu}_{\ch{O2}/\ch{FeO}}=\nu_{\ch{O2}/\ch{FeO}} \rho_{\ch{FeO}}  k_{0,\ch{FeO}}$ and $\tilde{\nu}_{\ch{O2}/\ch{Fe3O4}}=\nu_{\ch{O2}/\ch{Fe3O4}} \rho_{\ch{Fe3O4}} k_{0,\ch{Fe3O4}}$.\\

\begin{figure}[h!]
\centerline{\includegraphics[width=1.0\textwidth]{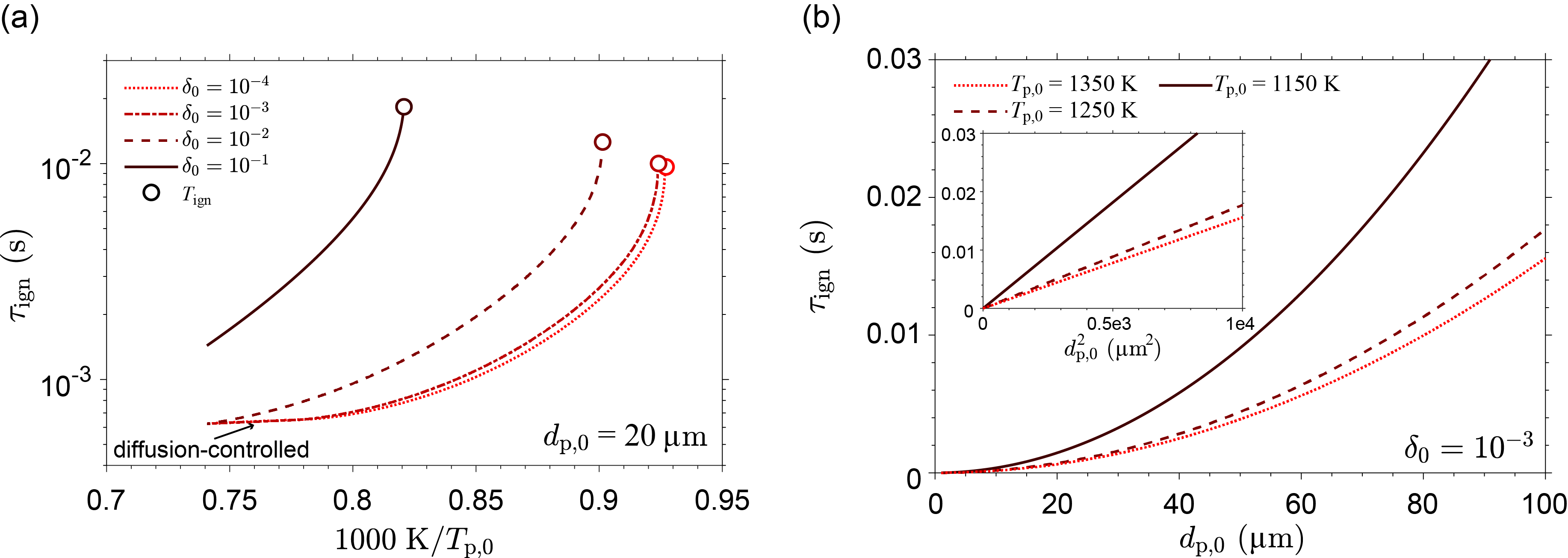}}
		\caption{The model prediction of $\tau_\mathrm{ign}$ as (a) a function of the reciprocal of initial particle temperature, $T_\mathrm{p,0}$, with $d_\mathrm{p,0}=\SI{20}{\micro\metre}$ and various values of $\delta_0$, and (b) a function of initial particle size, $d_\mathrm{p,0}$, with $\delta_0=10^{-3}$ and various values of $T_\mathrm{p,0}$. The inset of (b) shows the plot of $\tau_\mathrm{ign}$ as a function of $d^{2}_\mathrm{p,0}$.}
	\label{FigTauIgnCurves}
\end{figure}

As shown in Fig.~\ref{FigTauIgn}(b), below the boundary of the completely diffusion-controlled regime at \SI{1281}{\kelvin}, the iso-contours of $\tau_\mathrm{ign}$ (i.e., the dashed curves) show that a larger particle requires a greater initial temperature to result in the same ignition delay time. In the completely diffusion-controlled regime, the iso-contours of $\tau_\mathrm{ign}$ are nearly vertical, indicating that, for a fixed initial particle size, $\tau_\mathrm{ign}$ very slightly depends on particle temperature. This slight dependence is owing to the temperature-dependence of the transport properties of the gaseous species near the particle surface. The contour plot in Fig.~\ref{FigTauIgn}(b) shows that $\tau_\mathrm{ign}$ significantly increases with initial particle size. This trend is further clarified in Fig.~\ref{FigTauIgnCurves}(b) showing $\tau_\mathrm{ign}$ as functions of $d_\mathrm{p,0}$ for $\delta_0=10^{-3}$ and different values of $T_\mathrm{p,0}$. The inset of Fig.~\ref{FigTauIgnCurves}(b) demonstrates that $\tau_\mathrm{ign}$ scales with $d^{2}_\mathrm{p,0}$ at all of the values of $T_\mathrm{p,0}$ corresponding to both kinetic- and diffusion-controlled regimes. The ignition delay time is proportional to the thermal capacity of the particle and inversely proportional to the reaction rate. As shown in Eq.~\ref{EqmRmD}, both kinetic- and diffusion-controlled reaction rates linearly scale with particle size. The thermal capacity of the particle scales with particle volume, i.e., $\propto \; d^{3}_\mathrm{p,0}$. Thus, the scaling between $\tau_\mathrm{ign}$ and particle size follows a $d^2$-law\footnote{It is of interest to note that, if the oxidation kinetics follow a linear rate law (e.g., in the generic particle combustion model proposed by Soo~\textit{et al}.~\cite{Soo2018}), $\dot{m}_\mathrm{R}$ scales with particle surface area, i.e., $\propto \; d^{2}_\mathrm{p,0}$, and thus, the scaling of $\tau_\mathrm{ign}$ with particle size must follow a $d$-law.}.\\

\subsubsection{\label{Sect5_7_3}Ignition delay time of adiabatic suspensions of particles}

To better examine the dependence of $\tau_\mathrm{ign}$ on the reaction nature of iron particles, the model prediction of $\tau_\mathrm{ign}$ for adiabatic suspensions is analyzed in this section. As the size of a spherical suspension of particles becomes sufficiently large, i.e., $r_\mathrm{s} \; \to \; \infty$, the conductive heat transfer from the boundary of the suspension to the inert surrounding becomes negligible. An adiabatic suspension can thus be understood as an infinitely large cloud of suspended particles. The critical ignition temperature for such a suspension goes to the limit of zero. In other words, any initial temperature can trigger a thermal runaway of an adiabatic suspension. Thus, $\tau_\mathrm{ign}$, consistently defined as the time at which the particle temperature reaches the melting point of \ch{FeO}, can be measured over a broad range of initial particle (and gas) temperatures.\\

\begin{figure}[h!]
\centerline{\includegraphics[width=0.6\textwidth]{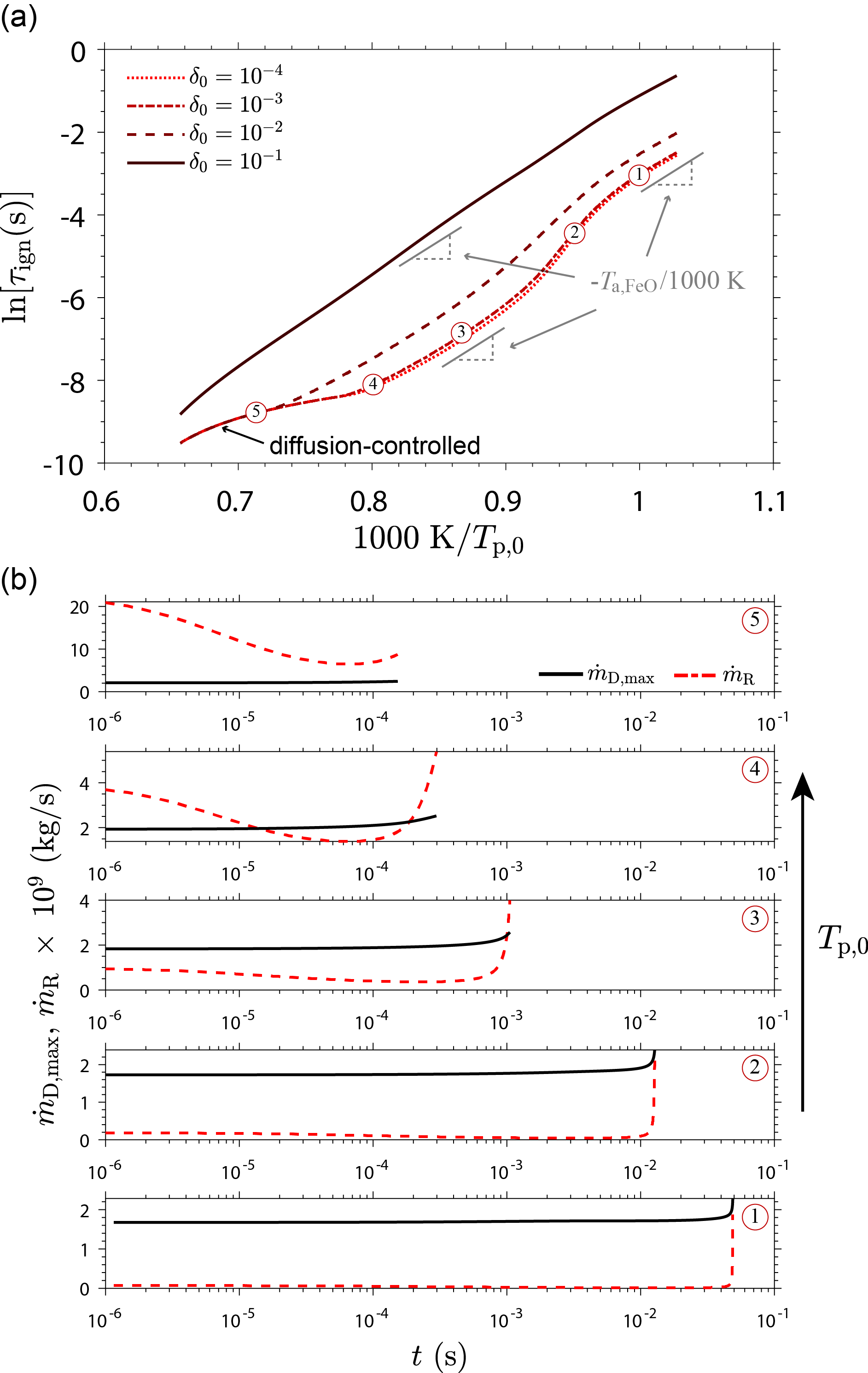}}
		\caption{(a) The natural logarithm of $\tau_\mathrm{ign}$ predicted for adiabatic suspensions of iron particles with $d_\mathrm{p,0}=\SI{20}{\micro\metre}$ and various values of $\delta_0$, plotted as functions of the reciprocal of initial particle temperature, i.e., $\SI{1000}{\kelvin}/T_\mathrm{p,0}$. (b) Comparison between the kinetic consumption rate, $\dot{m}_\mathrm{R}$, and the maximum external diffusion rate, $\dot{m}_\mathrm{D,max}$, of \ch{O2} in air for the cases with $\delta_0=10^{-3}$ and \textcircled{\small{1}} $T_\mathrm{p,0}=\SI{1000}{\kelvin}$, \textcircled{\small{2}} $T_\mathrm{p,0}=\SI{1050}{\kelvin}$, \textcircled{\small{3}} $T_\mathrm{p,0}=\SI{1150}{\kelvin}$, \textcircled{\small{4}} $T_\mathrm{p,0}=\SI{1250}{\kelvin}$, and \textcircled{\small{5}} $T_\mathrm{p,0}=\SI{1400}{\kelvin}$, as marked in (a).}
	\label{FigTauIgnSuspension}
\end{figure}

Figure~\ref{FigTauIgnSuspension}(a) shows the natural logarithm of $\tau_\mathrm{ign}$ predicted for adiabatic suspensions of iron particles with $d_\mathrm{p,0}=\SI{20}{\micro\metre}$ and various values of $\delta_0$, plotted as functions of the reciprocal of initial particle temperature, i.e., $\SI{1000}{\kelvin}/T_\mathrm{p,0}$. Over the entire range of temperature from \SI{973}{\kelvin} to \SI{1523}{\kelvin}, the predicted curve for the case with a relatively thick initial oxide layer of $\delta_0=10^{-1}$ follows a nearly linear trend with a slope of the activation temperature of \ch{FeO} formation. This linear trend indicates that the reaction rate of the particles is kinetic-controlled during the ignition process. As $\delta_0$ decreases from $10^{-1}$ to $10^{-4}$, the curves of $\ln{(\tau_\mathrm{ign})}$ vs. $1/T_\mathrm{p,0}$ become increasingly nonlinear.\\ 

To understand this nonlinear trend, the kinetic rate of oxygen consumption, $\dot{m}_\mathrm{R}$, and the maximum external-diffusion rate, $\dot{m}_\mathrm{D,max}$, for the cases with $\delta_0=10^{-3}$ and five selected values of $T_\mathrm{p,0}$ (as marked in Fig.~\ref{FigTauIgnSuspension}(a)) across the herein studied temperature range, are compared in Fig.~\ref{FigTauIgnSuspension}(b). At a sufficiently high temperature, e.g., Case \textcircled{\small{5}} with $T_\mathrm{p,0}=\SI{1400}{\kelvin}$, the ignition process is completely diffusion-controlled, as $\dot{m}_\mathrm{D,max}<\dot{m}_\mathrm{R}$ throughout the entire process. In this completely diffusion-controlled regime, $\tau_\mathrm{ign}$ becomes independent of oxide layer thickness, as shown in Fig.~\ref{FigTauIgnSuspension}(a) that the curves of $10^{-2}$, $10^{-3}$, and $10^{-4}$ collapse. For Case \textcircled{\small{4}} with $T_\mathrm{p,0}=\SI{1250}{\kelvin}$, the ignition process is partly diffusion-controlled as indicated by the double-transition behavior identified in Fig.~\ref{FigTauIgnSuspension}(b). The gradual increase in the slope of the $\ln{(\tau_\mathrm{ign})}$-vs.-$1/T_\mathrm{p,0}$ curve from Case \textcircled{\small{5}} to Case \textcircled{\small{4}}, as shown in Fig.~\ref{FigTauIgnSuspension}(a), can be attributed to this partly diffusion-controlled regime. Cases \textcircled{\small{1}}-\textcircled{\small{3}} with relatively low initial temperatures undergo a completely kinetic-controlled ignition process, as shown by the comparison between $\dot{m}_\mathrm{D,max}$ and $\dot{m}_\mathrm{R}$ plotted in Fig.~\ref{FigTauIgnSuspension}(b). The resulting $\ln{(\tau_\mathrm{ign})}$-vs.-$1/T_\mathrm{p,0}$ curve follows a nearly linear trend with a slope of $-T_\mathrm{a,FeO}$ over two separated ranges of $T_\mathrm{p,0}$, which are marked by Cases \textcircled{\small{1}} and \textcircled{\small{3}}. In the intermediate temperature range marked by Case \textcircled{\small{2}}, $\tau_\mathrm{ign}$ decreases more rapidly with an increase in $T_\mathrm{p,0}$ than the slope of $-T_\mathrm{a,FeO}$. This anomalous dependence of ignition delay time on temperature is likely due to the fact that $\tau_\mathrm{ign}$ is measured as the time when $T_\mathrm{p}$ reaches a fixed temperature, i.e., the melting point of \ch{FeO}. Thus, the particles with a higher initial temperature, not only have a greater kinetic reaction rate, but also require a smaller amount of energy release to reach the melting point of \ch{FeO}.\\

\subsubsection{\label{Sect5_7_3}Ignition delay time of non-adiabatic suspensions of particles}

\begin{figure}[h!]
\centerline{\includegraphics[width=0.6\textwidth]{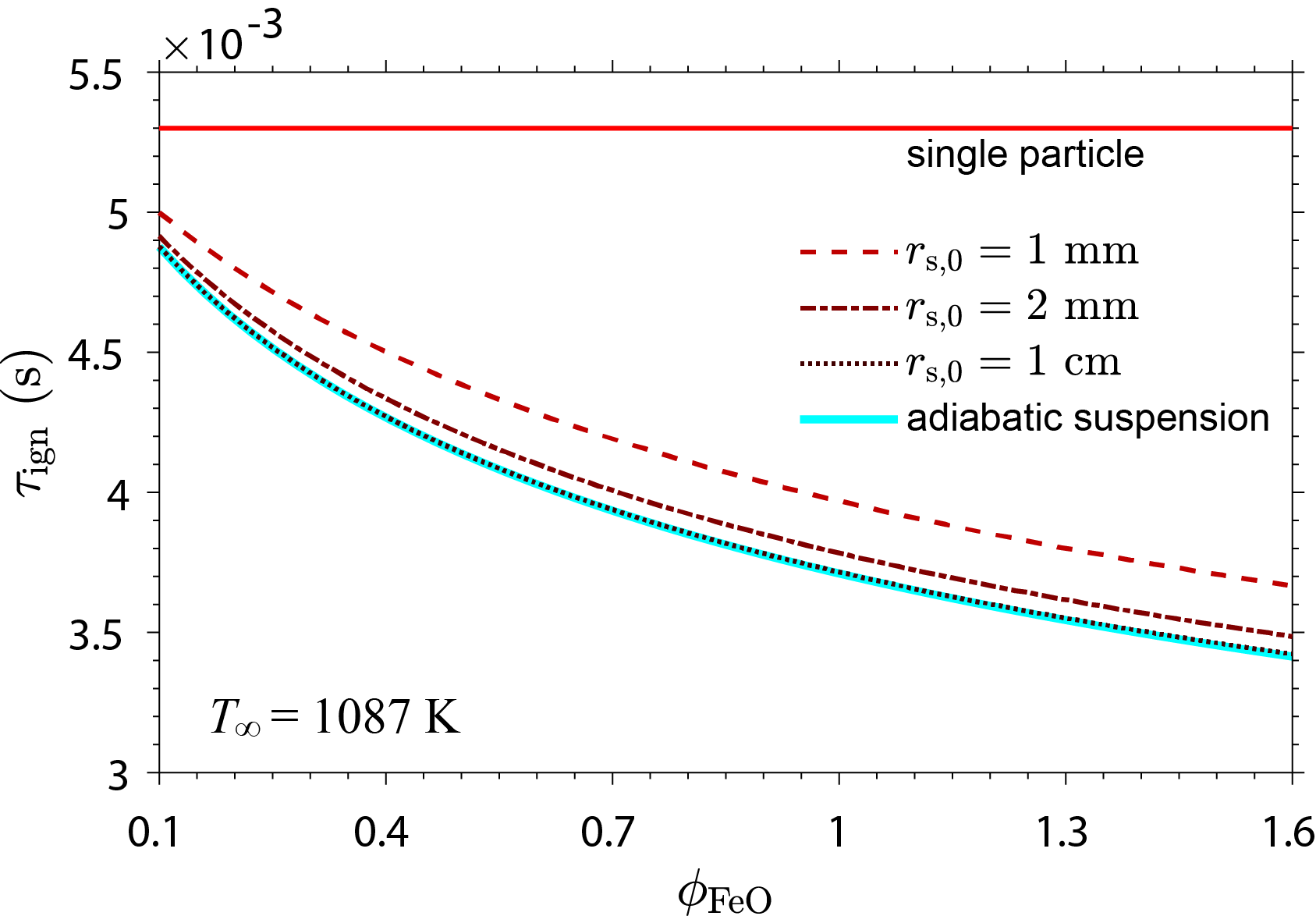}}
		\caption{The ignition delay time, $\tau_\mathrm{ign}$, of a non-adiabatic suspension of iron particles in air with different initial suspension radii and $\delta_0=10^{-3}$, as a function of fuel equivalence ratio (assuming \ch{FeO} as the only product) with $d_\mathrm{p,0}=\SI{20}{\micro\meter}$ at a fixed temperature \SI{1087}{\kelvin}, i.e., slightly above $T_\mathrm{ign}$ for an isolated particle. The value of $\tau_\mathrm{ign}$ for an isolated particle in air is marked by the horizontal solid line. The $\tau_\mathrm{ign}$ for an adiabatic suspension is plotted as the thick cyan curve.}
	\label{FigTauIgnNonAdiabaticSuspension}
\end{figure}

To more clearly demonstrate the collective heating effect, the model prediction of $\tau_\mathrm{ign}$ as a function of $\phi_{\ch{FeO}}$ for non-adiabatic suspensions at a common surrounding gas temperature, \SI{1087}{\kelvin}, which is slightly above the $T_\mathrm{ign}$ for an isolated particle, is plotted in Fig.~\ref{FigTauIgnNonAdiabaticSuspension}. The resulting $\tau_\mathrm{ign}$ for suspensions of different initial sizes are on the same order of magnitude as $\tau_\mathrm{ign}$ for an isolated particle. For larger values of $\phi_{\ch{FeO}}$, $\tau_\mathrm{ign}$ of suspensions becomes increasingly lower than that for an isolated particle due to a more pronounced collective heating effect. Furthermore, for a sufficiently large non-adiabatic suspension, i.e., $r_\mathrm{s,0} \geq \SI{1}{\centi\meter}$, the resulting $\tau_\mathrm{ign}$ approaches to the values for adiabatic suspensions as indicated by the cyan curve in Fig.~\ref{FigTauIgnNonAdiabaticSuspension}. It has been demonstrated in Sect.~\ref{Sect5_5} that the collective heating mechanism can significantly reduce the ignition temperature of sufficiently large non-adiabatic suspensions with sufficiently high particle number densities. Note that, at a reduced $T_\mathrm{ign}$, the ignition delay time of a suspension can be significantly longer than that of an isolated particle at its corresponding, higher, $T_\mathrm{ign}$.\\

\subsection{\label{Sect5_6}Comparison with experimental measurement}

The ignition temperatures measured by Grosse and Conway~\cite{Grosse1958} and Bolobov~\cite{Bolobov2001} for \SI{}{\milli\meter}-sized iron specimens are perhaps the closest to the definition of $T_\mathrm{ign}$ in the current paper. In these authors' experiments, a specimen was first heated in vacuum or an inert gas to a steady and uniform temperature $T_0$; oxygen or air was then fed into the test chamber; the minimum $T_0$ triggering an abrupt increase in the temperature of the specimen, i.e., a thermal runaway, was determined as the critical temperature for ignition. The ignition temperatures reported by Grosse and Conway~\cite{Grosse1958} and Bolobov~\cite{Bolobov2001} are $1203 \pm 10 \mathrm{K}$ and $1233 \pm 20 \mathrm{K}$, respectively. These values are close, but slightly greater, than the range of $T_\mathrm{ign}$ predicted by the current model for an isolated iron particle in air, i.e., from \SI{1080}{\kelvin} to \SI{1220}{\kelvin} for $\delta_0=10^{-4}$-$10^{-1}$ (as shown in Fig~\ref{Fig7}). This small discrepancy can be attributed to the following reasons: (1) The iron specimens in these experiments were in direct contact with the supporting device, thus, losing heat to the surrounding via solid-phase thermal conduction; (2) a \SI{}{\milli\meter}-sized specimen might be subjected to a natural convection with Rayleigh number greater than unity (as considered in the analysis of Bolobov and Podlevskikh~\cite{BolobovPodlevskikh2001}); (3) the supplied oxygen flow might be of a non-negligible velocity at a lower temperature than the specimen temperature.\\

Bolobov~\cite{Bolobov2001} also found that the ignition temperature of iron is barely dependent on oxygen concentration over a broad range of $p_{\ch{O2}}$ from $0.2$ to \SI{20}{\mega\pascal}. This finding is consistent with the fact that the growth rate of the \ch{FeO} and \ch{Fe3O4} layers is controlled by the lattice diffusion of \ch{Fe} ions subjected to the equilibrium \ch{Fe} activities at the interlayer boundaries, and not significantly affected by the ambient \ch{O2} concentration. Thus, the consideration of \ch{O2}-concentration-independent kinetic rates in the current analysis is supported by this experimental evidence.\\

As reviewed by Gorokhov~\cite{Gorokhov1964}, the experimentally measured ignition temperatures for clouds of sub-\SI{}{\milli\meter}-sized iron particles scatter over a range from \SI{588}{\kelvin} to \SI{1053}{\kelvin}. This significant reduction in ignition temperature from those measured for isolated iron specimens~\cite{Grosse1958,Bolobov1991,Bolobov2001} is likely due to the collective effect as captured by the current suspension model and discussed in Sect.~\ref{Sect5_4}. However, it is of importance to note that, since the values of the kinetic parameters (in Table~\ref{Tab1}) are only valid for iron oxidation over a temperature range from \SI{973}{\kelvin} to \SI{1523}{\kelvin}, the current model prediction below this range might not be quantitatively accurate.\\

\begin{table}[H]
\begin{center}
\caption{Comparison between experimental measurements and predictions by the current model of $T_\mathrm{ign}$ of various iron (or steel) specimens.}
\label{Tab3}
\begin{tabular}{| P{0.3\linewidth} | P{0.1\linewidth} | P{0.25\linewidth} | P{0.2\linewidth} |}
\hline
 & $T_\mathrm{ign}$ (K) & Specimen & Conditions\\
\hline
Grosse \& Conway, exp.~\cite{Grosse1958} & $1203 \pm 10$ & 10-g-weighted iron & Flow of \ch{O2} at 1 atm\\
\hline
Bolobov, exp.~\cite{Bolobov2001} & $1233 \pm 20$ & $5 \times 5 \times \SI{0.5}{\milli\meter}$ low-carbon steel foil & \ch{O2} at 0.14-0.6~MPa\\
\hline
Gorokhov (reviewed), exp.~\cite{Gorokhov1964} & $588$-$1053$ & Cloud of sub-mm Fe particles & Air at 1 atm\\
\hline
Gorokhov (reviewed), exp.~\cite{Gorokhov1964} & $553$-$733$ & Precipitated layer of sub-mm Fe particles & Air at 1 atm\\
\hline
Current analysis & $1080$ & Isolated Fe particle with $\delta_0 \lesssim 0.003$ & Air at 1 atm\\
\hline
Current analysis & $760$-$902$ & Spherical suspensions of Fe particles of $d_\mathrm{p,0}=\SI{20}{\micro\meter}$, $\phi_{\ch{FeO}}=1$, and $r_\mathrm{s,0}=\SI{5}{\centi\meter}$-$\SI{1}{\centi\meter}$ & Air at 1 atm\\
\hline
\end{tabular}
\end{center} 
\end{table}
Another possible source of inaccuracy is rooted in the assumption of spherical particles made in this study. Pulverized sponge iron (PSI) particles, e.g., those used for combustion experiments by T\'{o}th \textit{et al}.~\cite{Toth2020}, have irregular shapes and porous structures. Iron particles produced by the reduction of spherical iron-oxide particles are porous because the iron oxides take up more volume than iron due to the fact that the Pilling-Bedworth ratios for $\ch{Fe} \to \ch{Fe2O3}$, \ch{Fe3O4}, and \ch{FeO} are all greater than unity. An irregularly shaped, porous iron particle has a greater surface area that has access to and can react with oxygen. Thus, the ignition temperature might be significantly reduced. Future efforts are required to quantitatively measure the porosity of reduced iron particles and estimate how particle porosity enhances the ignition of particles and their suspensions.\\

It is noteworthy that, in reality, the oxidation rate of iron is governed by different processes when the oxide layer is extremely thin, i.e., under approximately \SI{20}{\nano\meter}. The parabolic rate law considered in this analysis is based upon the assumption that Fe-cation diffusion across a sufficiently thick oxide layer with a uniform electric field is the rate-limiting process.  However, for a very thin oxide layer, its growth is governed by ionic transfer driven by a large electric field and space charges in the oxide layer~\cite{Atkinson1985}. In this thin-layer regime, the oxidation process is better explained by the theory of Cabrera and Mott~\cite{CabreraMott1949}. As a quantitatively reliable rate law for thin-layer growth of iron oxidation is not available, the thick-layer parabolic rate law is extrapolated in the current analysis to a thin-layer regime. The question arises as to how this extrapolation affects the accuracy in estimating the ignition characteristics of iron particles. An example of the growth rate of \ch{NiO} as a function of oxide layer thickness in both thin- and thick-layer regimes is provided in Fig.~7 of Atkinson's review~\cite{Atkinson1985}: If one extrapolates the parabolic rate law (Wagner's model) to nano-scale thicknesses, the growth rate would be underestimated compared to the prediction of Cabrera and Mott's model. An underestimation of the initial oxide-layer growth rate in the current analysis for iron is unlikely to have a significant effect on $T_\mathrm{ign}$. As demonstrated in Fig.~\ref{Fig8}(b) and (d) and discussed in Sect.~\ref{Sect5_2}, for a sufficiently thin initial oxide layer, the initial rapid oxide-layer growth has no effect on the subsequent processes determining whether a thermal runaway occurs. Therefore, the thin layer mechanisms described by the theory of Cabrera and Mott, which result in greater oxidation rates over a time period much shorter than the thermal-runway time scale of an iron particle, are unlikely to have a significant effect on the ignition characteristics.\\

A recent study by Senyurt and Dreizin~\cite{SenyurtDreizin2022} demonstrates that, if the size of a reactive metal particle is comparable to the molecular mean free path of the surrounding gas, the transport processes between the particle and the gas are hindered by the rarefaction effects in the Knudsen transition regime. These authors' analysis shows that, owing to a reduced rate of heat loss to the surrounding in the Knudsen transition regime, the ignition temperatures of sufficiently small particles are significantly lower than the model prediction based upon a continuum assumption. However, this analysis is limited to a steady-state model and not applied to study iron particles. A detailed, unsteady analysis of how the ignition characteristics of fine iron particles differ in the Knudsen transition regime will be reported in a future publication.\\

\section{\label{Sect6}Concluding remarks}

A model describing the heat and mass balance equations for heterogeneously burning micrometric iron particles was developed. The solid-phase oxidation kinetics described by a parabolic rate law were empirically calibrated using measured experimental data on the time-evolution of the growth of iron-oxide layers. The model was applied to quantitatively analyze the dependence of the minimum gas temperature required for a thermal runaway (namely, the ignition temperature) on particle size, initial thickness of the oxide layer, gas-phase composition, heat loss by radiation, and the collective heating effect in particle suspensions. Findings showed that the ignition temperature depends on the ratio between the initial oxide layer thickness and particle size, $\delta_0$, and is independent of particle size. For isolated iron particles in air, the predicted ignition temperature plateaus at approximately \SI{1080}{\kelvin}, and becomes independent of $\delta_0$, for $\delta_0 \lesssim 0.003$. The lower the thermal conductivity of the gas-phase environment, the lower the ignition temperature. The ignition behavior was found to be largely independent of radiative heat loss effects. The collective heating effect in burning particle suspensions was shown to significantly reduce the ignition temperature. The slight underestimation of the ignition temperature in the current model compared to experimental literature~\cite{Grosse1958,Bolobov2001}, is likely associated with experimental conditions that facilitated further heat loss from the iron specimens in addition to convective and radiative heat loss.\\

The time histories of the reaction rate of the particles were used to parametrically probe the transition behavior between kinetic-controlled and external-diffusion-controlled combustion regimes. The ignition delay time is significantly influenced by the initial oxide layer thickness and particle temperature under a kinetic-controlled combustion regime; once a particle transitions to an external-diffusion-controlled regime, its ignition delay time becomes nearly independent of $\delta_0$ and $T_\mathrm{p,0}$. Furthermore, the scaling of the ignition delay time of iron particles with particle size follows a $d^2$-law. Future experimental and modeling efforts should work to elucidate the rate of heat release in iron particle combustion in liquid-phase, and to explore how to ensure a non-volatile combustion of molten iron droplets.

\section*{Acknowledgement}
The authors are grateful to Y. Pyo for his contributions to the early stages of this study and to J. Pale\v{c}ka and S. Goroshin for useful discussion in developing this paper. X.C.M. thanks L. Thijs for pointing out some lacks of clarity in the model description. Funding was provided by the Canadian Space Agency through the Flights and Fieldwork for the Advancement of Science and Technology (FAST) Grant Program. A.F. was supported by a TISED Award for Summer Undergraduate Research (TASUR).

\bibliographystyle{ieeetr}
\bibliography{metalFlame}

\begin{thebibliography}{10}

\bibitem{Beach2006}
D.~Beach, A.~Rondinone, B.~Sumpter, S.~Labinov, and R.~Richards, ``{Solid-State
  Combustion of Metallic Nanoparticles: New Possibilities for an Alternative
  Energy Carrier},'' {\em Journal of Energy Resources Technology}, vol.~129,
  pp.~29--32, 07 2006.

\bibitem{Bergthroson2015Review}
J.~Bergthorson, S.~Goroshin, M.~Soo, P.~Julien, J.~Palecka, D.~Frost, and
  D.~Jarvis, ``Direct combustion of recyclable metal fuels for zero-carbon heat
  and power,'' {\em Applied Energy}, vol.~160, pp.~368 -- 382, 2015.

\bibitem{Bergthorson2018Review}
J.~Bergthorson, ``Recyclable metal fuels for clean and compact zero-carbon
  power,'' {\em Progress in Energy and Combustion Science}, vol.~68, pp.~169 --
  196, 2018.

\bibitem{Paidassi1958a}
J.~Pa\"{i}dassi, ``Sur la cinetique de l'oxydation du fer dans l'air dans
  l'intervalle $700$-${1250}^\circ \mathrm{C}$,'' {\em Acta Metallurgica},
  vol.~6, no.~3, pp.~184--194, 1958.

\bibitem{Paidassi1958b}
J.~Pa\"{i}dassi, ``Sur l'oxydation du protoxyde de fer dans l'air dans
  l'intervalle $600$-${1350}^\circ \mathrm{C}$,'' {\em Acta Metallurgica},
  vol.~6, no.~3, pp.~219--221, 1958.

\bibitem{Yurek1974}
G.~Yurek, J.~Hirth, and R.~Rapp, ``The formation of two-phase layered scales on
  pure metals,'' {\em Oxidation of Metals}, vol.~8, no.~5, pp.~265--281, 1974.

\bibitem{SmeltzerYoung1975}
W.~Smeltzer and D.~Young, ``Oxidation properties of transition metals,'' {\em
  Progress in Solid State Chemistry}, vol.~10, pp.~17--54, 1975.

\bibitem{GarnaudRapp1977}
G.~Garnaud and R.~Rapp, ``Thickness of the oxide layers formed during the
  oxidation of iron,'' {\em Oxidation of Metals}, vol.~11, no.~4, pp.~193--198,
  1977.

\bibitem{Sato1983}
J.~Sato, K.~Sato, and T.~Hirano, ``Fire spread mechanisms along steel cylinders
  in high pressure oxygen,'' {\em Combustion and Flame}, vol.~51, pp.~279--287,
  1983.

\bibitem{Hirano1993}
T.~Hirano and J.~Sato, ``Fire spread along structural metal pieces in oxygen,''
  {\em Journal of Loss Prevention in the Process Industries}, vol.~6, no.~3,
  pp.~151--157, 1993.

\bibitem{Steinberg1992}
T.~Steinberg, G.~Mulholland, D.~Wilson, and F.~Benz, ``The combustion of iron
  in high-pressure oxygen,'' {\em Combustion and Flame}, vol.~89, no.~2,
  pp.~221--228, 1992.

\bibitem{Steinberg1998}
T.~Steinberg, J.~Kurtz, and D.~Wilson, ``The solubility of oxygen in liquid
  iron oxide during the combustion of iron rods in high-pressure oxygen,'' {\em
  Combustion and Flame}, vol.~113, no.~1, pp.~27--37, 1998.

\bibitem{WardSteinberg2009}
N.~Ward and T.~Steinberg, ``The rate-limiting mechanism for the heterogeneous
  burning of cylindrical iron rods,'' {\em Journal of ASTM International},
  vol.~6, no.~6, pp.~1--13, 2009.

\bibitem{Muller2014CST}
M.~Muller, H.~El-Rabii, and R.~Fabbro, ``Laser ignition of bulk iron, mild
  steel, and stainless steel in oxygen atmospheres,'' {\em Combustion Science
  and Technology}, vol.~186, no.~7, pp.~953--974, 2014.

\bibitem{Muller2015}
M.~Muller, H.~El-Rabii, and R.~Fabbro, ``Liquid phase combustion of iron in an
  oxygen atmosphere,'' {\em Journal of Materials Science}, vol.~50, no.~9,
  pp.~3337--3350, 2015.

\bibitem{Wagner1933}
C.~Wagner, ``Beitrag zur theorie des anlaufvorgangs,'' {\em Zeitschrift f\"{u}r
  Physikalische Chemie}, vol.~21B, no.~1, pp.~25--41, 1933.

\bibitem{Hauffe1965}
K.~Hauffe, {\em The Mechanism of Oxidation of Metals---Theory}, pp.~79--143.
\newblock Boston, MA: Springer US, 1965.

\bibitem{Dreizin2000}
E.~Dreizin, ``Phase changes in metal combustion,'' {\em Progress in Energy and
  Combustion Science}, vol.~26, no.~1, pp.~57--78, 2000.

\bibitem{Hirano1983}
T.~Hirano, K.~Sato, Y.~Sato, and J.~Sato, ``Prediction of metal fire spread in
  high pressure oxygen,'' {\em Combustion Science and Technology}, vol.~32,
  no.~1-4, pp.~137--159, 1983.

\bibitem{ElRabii2017}
H.~El-Rabii, K.~Kazakov, and M.~Muller, ``Experimental and theoretical study of
  iron and mild steel combustion in oxygen flows,'' {\em Physics of Fluids},
  vol.~29, no.~3, p.~037104, 2017.

\bibitem{SunHirano2000}
J.~Sun, R.~Dobashi, and T.~Hirano, ``Combustion behavior of iron particles
  suspended in air,'' {\em Combustion Science and Technology}, vol.~150,
  no.~1-6, pp.~99--114, 2000.

\bibitem{Tang2011PROCI}
F.~Tang, S.~Goroshin, and A.~Higgins, ``Modes of particle combustion in iron
  dust flames,'' {\em Proceedings of the Combustion Institute}, vol.~33, no.~2,
  pp.~1975 -- 1982, 2011.

\bibitem{Julien2015}
P.~Julien, S.~Whiteley, S.~Goroshin, M.~Soo, D.~Frost, and J.~Bergthorson,
  ``Flame structure and particle-combustion regimes in premixed
  methane-iron-air suspensions,'' {\em Proceedings of the Combustion
  Institute}, vol.~35, no.~2, pp.~2431 -- 2438, 2015.

\bibitem{Wright2016}
A.~Wright, A.~Higgins, and S.~Goroshin, ``The discrete regime of flame
  propagation in metal particulate clouds,'' {\em Combustion Science and
  Technology}, vol.~188, no.~11-12, pp.~2178--2199, 2016.

\bibitem{McRae2018}
M.~McRae, P.~Julien, S.~Salvo, S.~Goroshin, D.~Frost, and J.~Bergthorson,
  ``Stabilized, flat iron flames on a hot counterflow burner,'' {\em
  Proceedings of the Combustion Institute}, 2018.

\bibitem{Palecka2019Perwaves}
J.~Pale{\v{c}}ka, J.~Sniatowsky, S.~Goroshin, A.~Higgins, and J.~Bergthorson,
  ``A new kind of flame: Observation of the discrete flame propagation regime
  in iron particle suspensions in microgravity,'' {\em Combustion and Flame},
  vol.~209, pp.~180--186, 2019.

\bibitem{Palecka2020Perwaves}
J.~Pale{\v{c}}ka, S.~Goroshin, A.~Higgins, Y.~Shoshin, P.~{de Goey}, J.-R.
  Angilella, H.~Oltmann, A.~Stein, B.~Schmitz, A.~Verga, S.~Vincent-Bonnieu,
  W.~Sillekens, and J.~Bergthorson, ``Percolating reaction–diffusion waves
  (perwaves)—sounding rocket combustion experiments,'' {\em Acta
  Astronautica}, vol.~177, pp.~639--651, 2020.

\bibitem{Poletaev2020}
N.~Poletaev and M.~Khlebnikova, ``Combustion of iron particles suspension in
  laminar premixed and diffusion flames,'' {\em Combustion Science and
  Technology}, vol.~0, no.~0, pp.~1--22, 2020.

\bibitem{Tang2009CTM}
F.~Tang, A.~Higgins, and S.~Goroshin, ``Effect of discreteness on heterogeneous
  flames: propagation limits in regular and random particle arrays,'' {\em
  Combustion Theory and Modelling}, vol.~13, no.~2, pp.~319--341, 2009.

\bibitem{Tang2011PRE}
F.~Tang, A.~Higgins, and S.~Goroshin, ``Propagation limits and velocity of
  reaction-diffusion fronts in a system of discrete random sources,'' {\em
  Physical Review E}, vol.~85, p.~036311, Mar 2012.

\bibitem{Goroshin2011PRE}
S.~Goroshin, F.~Tang, and A.~Higgins, ``Reaction-diffusion fronts in media with
  spatially discrete sources,'' {\em Physical Review E}, vol.~84, no.~2,
  p.~027301, 2011.

\bibitem{Mi2016PROCI}
X.~Mi, A.~Higgins, S.~Goroshin, and J.~Bergthorson, ``The influence of spatial
  discreteness on the thermo-diffusive instability of flame propagation with
  infinite lewis number,'' {\em Proceedings of the Combustion Institute},
  vol.~36, no.~2, pp.~2359--2366, 2017.

\bibitem{Lam2017PRE}
F.~Lam, X.~Mi, and A.~Higgins, ``Front roughening of flames in discrete
  media,'' {\em Phys. Rev. E}, vol.~96, p.~013107, Jul 2017.

\bibitem{Lam2018CTM}
F.~Lam, X.~Mi, and A.~Higgins, ``Dimensional scaling of flame propagation in
  discrete particulate clouds,'' {\em Combustion Theory and Modelling}, vol.~0,
  no.~0, pp.~1--24, 2019.

\bibitem{Toth2020}
P.~T\'oth, Y.~\"Ogren, A.~Sepman, P.~Gren, and H.~Wiinikka, ``Combustion
  behavior of pulverized sponge iron as a recyclable electrofuel,'' {\em Powder
  Technology}, vol.~373, pp.~210--219, 2020.

\bibitem{Ning2021}
D.~Ning, Y.~Shoshin, J.~{van Oijen}, G.~Finotello, and L.~{de Goey}, ``Burn
  time and combustion regime of laser-ignited single iron particle,'' {\em
  Combustion and Flame}, vol.~230, p.~111424, 2021.

\bibitem{Huang2021}
J.~Huang, S.~Li, W.~Cai, Y.~Qian, E.~Berrocal, M.~Ald\'{e}n, and Z.~Li,
  ``Quantification of the size, 3d location and velocity of burning iron
  particles in premixed methane flames using high-speed digital in-line
  holography,'' {\em Combustion and Flame}, vol.~230, p.~111430, 2021.

\bibitem{Ning2022CNF}
D.~Ning, Y.~Shoshin, M.~{van Stiphout}, J.~{van Oijen}, G.~Finotello, and
  P.~{de Goey}, ``Temperature and phase transitions of laser-ignited single
  iron particle,'' {\em Combust. Flame}, vol.~236, p.~111801, 2022.

\bibitem{Gorokhov1964}
Y.~Gorokhov, ``On the pyrophoric properties, explosion hazard, and toxicity of
  powders and dusts of iron and its compounds,'' {\em Soviet Powder Metallurgy
  and Metal Ceramics}, vol.~3, no.~1, pp.~82--86, 1964.

\bibitem{Leshchevich2012}
V.~Leshchevich, O.~Penyazkov, A.~Fedorov, A.~Shul'gin, and J.~Rostaing,
  ``Conditions and delay time of ignition of iron microparticles in oxygen,''
  {\em Journal of Engineering Physics and Thermophysics}, vol.~85, no.~1,
  pp.~148--154, 2012.

\bibitem{Grosse1958}
A.~Grosse and J.~Conway, ``Combustion of metals in oxygen,'' {\em Industrial \&
  Engineering Chemistry}, vol.~50, no.~4, pp.~663--672, 1958.

\bibitem{Bolobov1991}
V.~Bolobov, A.~Berezin, P.~Drozhzhin, and A.~Shteinberg, ``Ignition of compact
  stainless steel specimens in high pressure oxygen,'' {\em Combustion,
  Explosion and Shock Waves}, vol.~27, no.~3, pp.~263--266, 1991.

\bibitem{Bolobov2001}
V.~Bolobov, ``Conditions for ignition of iron and carbon steel in oxygen,''
  {\em Combustion, Explosion and Shock Waves}, vol.~37, no.~3, pp.~292--296,
  2001.

\bibitem{Breiter1977}
A.~Breiter, V.~Mal'tsev, and E.~Popov, ``Models of metal ignition,'' {\em
  Combustion, Explosion, and Shock Waves}, vol.~13, no.~4, pp.~475--485, 1977.

\bibitem{Khaikin1970Ignition}
B.~Khaikin, V.~Bloshenko, and A.~Merzhanov, ``On the ignition of metal
  particles,'' {\em Combustion, Explosion and Shock Waves}, vol.~6, no.~4,
  pp.~412--422, 1970.

\bibitem{BolobovPodlevskikh2001}
V.~Bolobov and N.~Podlevskikh, ``Numerical analysis of conditions for ignition
  of compact metal specimens and foil in oxygen,'' {\em Combustion, Explosion
  and Shock Waves}, vol.~37, no.~6, pp.~655--663, 2001.

\bibitem{ChenYuen2003}
R.~Chen and W.~Yeun, ``Review of the high-temperature oxidation of iron and
  carbon steels in air or oxygen,'' {\em Oxidation of metals}, vol.~59, no.~5,
  pp.~433--468, 2003.

\bibitem{Li2011EnergyEnviron}
F.~Li, Z.~Sun, S.~Luo, and L.~Fan, ``Ionic diffusion in the oxidation of
  iron—effect of support and its implications to chemical looping
  applications,'' {\em Energy Environ. Sci.}, vol.~4, pp.~876--880, 2011.

\bibitem{Himmel1953JOM}
L.~Himmel, R.~Mehl, and C.~Birchenall, ``Self-diffusion of iron in iron oxides
  and the {Wagner} theory of oxidation,'' {\em JOM}, vol.~5, no.~6,
  pp.~827--843, 1953.

\bibitem{Goursat1973kinetics}
A.~Goursat and W.~Smeltzer, ``Kinetics and morphological development of the
  oxide scale on iron at high temperatures in oxygen at low pressure,'' {\em
  Oxidation of Metals}, vol.~6, no.~2, pp.~101--116, 1973.

\bibitem{Young2008}
D.~Young, {\em High temperature oxidation and corrosion of metals}, vol.~1.
\newblock Elsevier, 2008.

\bibitem{Kofstad1972}
P.~Kofstad, ``Nonstoichiometry, diffusion, and electrical conductivity in
  binary metal oxides,'' 1972.

\bibitem{Atkinson1985}
A.~Atkinson, ``Transport processes during the growth of oxide films at elevated
  temperature,'' {\em Rev. Mod. Phys.}, vol.~57, pp.~437--470, Apr 1985.

\bibitem{Carter1959}
R.~Carter, ``Thermal expansion of mgfe2o4, feo, and mgo{\textperiodcentered}
  2feo,'' {\em Journal of the American Ceramic Society}, vol.~42, no.~7,
  pp.~324--327, 1959.

\bibitem{Holcomb2019}
G.~Holcomb, ``A review of the thermal expansion of magnetite,'' {\em Materials
  at High Temperatures}, vol.~36, no.~3, pp.~232--239, 2019.

\bibitem{Kozlovskii2019}
Y.~Kozlovskii and S.~Stankus, ``The linear thermal expansion coefficient of
  iron in the temperature range of 130--1180 k,'' in {\em Journal of Physics:
  Conference Series}, vol.~1382, p.~012181, IOP Publishing, 2019.

\bibitem{Chase_NIST}
M.~Chase, {\em {NIST}-{JANAF} Thermochemical Tables, 4th Edition}.
\newblock American Institute of Physics, -1, 1998-08-01 1998.

\bibitem{Mcbride1993}
B.~McBride, {\em Coefficients for calculating thermodynamic and transport
  properties of individual species}, vol.~4513.
\newblock NASA Langley Research Center, 1993.

\bibitem{Hubbard1975}
G.~Hubbard, V.~Denny, and A.~Mills, ``Droplet evaporation: Effects of
  transients and variable properties,'' {\em International Journal of Heat and
  Mass Transfer}, vol.~18, no.~9, pp.~1003--1008, 1975.

\bibitem{Hazenberg2021}
T.~Hazenberg and J.~{van Oijen}, ``Structures and burning velocities of flames
  in iron aerosols,'' {\em Proceedings of the Combustion Institute}, vol.~38,
  no.~3, pp.~4383--4390, 2021.

\bibitem{Burgess1915}
G.~Burgess and R.~Waltenberg, {\em The Emissivity of Metals and Oxides:
  Measurements with the micropyrometer}, vol.~11.
\newblock US Department of Commerce, Bureau of Standards, 1915.

\bibitem{Touloukian1972}
Y.~Touloukian and D.~DeWitt, ``Thermophysical properties of matter - the tprc
  data series. volume 8. thermal radiative properties - nonmetallic solids,''
  Tech. Rep. {AD-A-951942/2/XAB}; {CNN: DSA900-73-C-2101}, Purdue University,
  Lafayette, IN, 1972.

\bibitem{Jones2019}
J.~Jones, P.~Mason, and A.~Williams, ``A compilation of data on the radiant
  emissivity of some materials at high temperatures,'' {\em Journal of the
  Energy Institute}, vol.~92, no.~3, pp.~523--534, 2019.

\bibitem{Soo2018}
M.~Soo, X.~Mi, S.~Goroshin, A.~Higgins, and J.~Bergthorson, ``Combustion of
  particles, agglomerates, and suspensions - a basic thermophysical analysis,''
  {\em Combustion and Flame}, vol.~192, pp.~384 -- 400, 2018.

\bibitem{Mi2013}
X.~Mi, S.~Goroshin, A.~Higgins, R.~Stowe, and S.~Ringuette, ``Dual-stage
  ignition of boron particle agglomerates,'' {\em Combustion and Flame},
  vol.~160, no.~11, pp.~2608--2618, 2013.

\bibitem{CabreraMott1949}
N.~Cabrera and N.~Mott, ``Theory of the oxidation of metals,'' {\em Reports on
  Progress in Physics}, vol.~12, pp.~163--184, jan 1949.

\bibitem{SenyurtDreizin2022}
E.~Senyurt and E.~Dreizin, ``At what ambient temperature can thermal runaway of
  a burning metal particle occur?,'' {\em Combust. Flame}, vol.~236, p.~111800,
  2022.

\end{thebibliography}

\end{document}